\documentclass{article}
\usepackage[top=3cm,right=2.3cm,bottom=3cm,left=2.3cm]{geometry}

\usepackage[scr=euler]{mathalfa}

\usepackage{graphicx} 
\usepackage{tikz}
\usetikzlibrary{matrix,positioning,arrows.meta,shapes}
\usepackage{caption}
\usepackage[caption = false]{subfig}

\usepackage{diagbox}
\usepackage{booktabs}
\usepackage{enumitem}
\usepackage{threeparttable}

\usepackage{amsmath, amssymb}
\usepackage{resizegather} 
\usepackage{dsfont}

\usepackage{bm}

\usepackage{natbib}
\usepackage{multirow}
\usepackage{url}

\usepackage{algorithm}
\usepackage{algpseudocode}

\usepackage{float}

\title{Heterogeneous behavioral mechanisms in epidemiological models}

\author{Jessica Pavani\textsuperscript{1}, Rob Deardon\textsuperscript{1,2} \& Alexandra M. Schmidt\textsuperscript{3} \\
\textsuperscript{1}\small Department of Mathematics and Statistics, University of Calgary, Calgary, AB, Canada \\
\textsuperscript{2}\small Faculty of Veterinary Medicine, University of Calgary, Calgary, AB, Canada \\
\textsuperscript{3}\small Department of Epidemiology, Biostatistics, and Occupational Health, McGill University, Montreal, QC, Canada
}

\date{}

\begin{document}

\maketitle

\begin{abstract}
Traditional epidemic models frequently assume behavioral homogeneity. The susceptible-infected-recovered model provides a robust foundation for characterizing disease transmission, but it does so without accounting for how people actually respond to risk. In contrast, behavioral‑change models incorporate mechanisms that capture how individuals adjust their actions during an outbreak, recognizing that rising infection risk typically motivates protective behaviors. Yet both approaches share a key limitation: they overlook the inherent heterogeneity of a population. In reality, communities are a complex mixture of risk tolerances and behavioral tendencies. Ignoring this inherent heterogeneity can obscure important differences in how individuals perceive and respond to disease threats. This paper introduces a novel Bayesian mixture model designed to address this limitation by partitioning the population into two distinct behavioral patterns: risk-neutral individuals, who maintain baseline contact rates, and risk-averse individuals, who modulate their behavior in response to epidemic severity. By integrating these disparate dynamics into a unified transmission framework, the proposed model explicitly accounts for varying population behaviors often overlooked by aggregate approaches. Through simulation studies and empirical data applications, we demonstrate that this mixture approach significantly outperforms traditional models in parameter recovery, epidemic trajectory estimation, and forecasting precision. Our findings suggest that failing to account for behavioral diversity leads to biased peak estimates and artificially stretched epidemic curves. Consequently, this research provides a more nuanced computational toolkit for predicting outbreak trajectories in socially fragmented environments, ensuring that public health intervention strategies are informed by a foundation of behavioral realism.

\vspace{0.1cm}

\noindent \textbf{Keywords}: Bayesian inference, Compartmental model, COVID-19, Infectious diseases, SIR. 
\end{abstract}


\section{Introduction} \label{sec:intro}

The mathematical modeling of infectious diseases has become an indispensable component of public health infrastructure, providing critical foresight into the transmission dynamics of global outbreaks. As pathogens emerge and spread, the ability to accurately predict epidemic trajectories is paramount for the strategic allocation of medical resources and the effective implementation of non‑pharmaceutical interventions. Since its foundation in the early twentieth century by pioneers such as Hamer, Ross, and most notably \citet{Kermack1927}, the susceptible-infected-recovered (SIR) model has served as the definitive cornerstone of mathematical epidemiology. This compartmental framework elegantly describes the transition of individuals through infection stages using a system of coupled equations, either deterministically through nonlinear ordinary differential equations or stochastically, with statistical inference typically carried out under a Bayesian paradigm. Despite its foundational assumption of population homogeneity and the absence of an explicit closed‑form solution, the model’s mathematical tractability allows researchers to extract profound insights into outbreak dynamics. Its enduring utility is evidenced by a vast body of literature that has extended this basic architecture in increasingly sophisticated ways, enabling the study of more complex disease‑transmission scenarios through the introduction of additional compartments and more intricate flows between them \citep{Satsuma2004, Xu2016, Agaba2017, Cooper2020, Gu2022, Lazebnik2023}. 

Although classical compartmental models remain the primary structure upon which modern epidemic modeling is built, they often lack the flexibility to reflect real‑world transmission dynamics, as the transmission rate is typically assumed to remain fixed. This limitation has motivated the incorporation of human behavior into compartmental modeling. Behavioral Change (BC) models represent a critical evolution in this field, replacing static transmission parameters with dynamic mechanisms modulated by a perceived alarm level, which captures behavioral responses. As proposed by \cite{Ward2023}, the alarm level is represented by a function of recent disease case counts, such that its effect increases with recent incidence or prevalence, reflecting the uptake of protective measures in response to rising cases. When infections are absent, the alarm function reduces to zero, indicating no behavioral change. Recent studies have demonstrated the substantial utility of alarm functions in improving model fit and forecasting accuracy \citep{Ward2023, Ward2025, Ward2024}. Furthermore, by adopting a Bayesian paradigm, we can effectively learn about the entire process. Specifically, this formulation uses incidence‑based triggers to modify susceptibility and contact rates, enabling rigorous estimation of behavioral parameters via Markov Chain Monte Carlo (MCMC) methods. Consequently, BC models offer a superior alternative for public health modeling, providing the flexibility needed to detect subsequent incidence peaks and to account for the multifaceted nature of population‑level protective behaviors.

Despite these advancements, a significant limitation persists in the current epidemiological literature: the assumption of behavioral homogeneity. Most existing frameworks, including SIR and BC models, conceptualize the population as a monolithic entity, wherein every individual is assumed to adopt an identical alarm level and subsequent degree of risk reduction. However, empirical observations from recent global health crises suggest a far more fragmented social response, raising questions as to whether a single alarm function can accurately represent a population characterized by diverse socioeconomic constraints and varying levels of risk tolerance. Failing to account for this internal variance may lead to biased parameter estimation and inaccurate epidemic forecasts. To address this gap, the present study develops a framework that explicitly captures heterogeneity in population behavior. We propose a Bayesian mixture model that accommodates the coexistence of distinct behavioral groups within a single population. Specifically, our model incorporates a weight parameter that partitions the susceptible population into two latent subgroups: a risk-neutral cohort that maintains baseline contact rates, consistent with traditional SIR dynamics, and a risk-averse cohort that modifies its behavior according to the BC framework.

The remainder of this paper is organized as follows. First, Section~\ref{sec:back} reviews the traditional SIR and BC models that serve as the foundational elements for the proposed approach. Next, Section~\ref{sec:model} details the mathematical formulation of the mixture model, along with the Bayesian inference procedure implemented via MCMC methods. Then, Section~\ref{sec:simulations} presents the results of simulation studies evaluating model fit and forecasting performance. Subsequently, Section~\ref{sec:application} discusses the application of the model to COVID‑19 datasets from New York City and Montreal. Finally, concluding remarks are provided in Section~\ref{sec:discussion}.

\section{Background} \label{sec:back}

This section examines the key concepts necessary to understand the proposed model. In particular, it reviews the classical SIR and BC models, which serve as the foundational elements for the development process.

\subsection{Susceptible-infected-recovered model} \label{subsec:SIR}

According to the SIR model introduced by \cite{Kermack1927}, individuals initially belong to the susceptible compartment and transition to the infectious and recovered compartments upon infection. The model presumes a homogeneous population, meaning that every individual has an equal probability of interacting with others and transmitting the disease at any given point in time. Different from the original proposal, this study describes the dynamics of compartment membership over time using a series of discrete‑time difference equations, which are well suited to analyzing regularly observed data. To formalize this, we define discrete time units $t = 1, \ldots, T$ representing specific times on the calendar \citep{Bjoernstad2002}. At each time $t$, the variables $S_{t}$, $I_{t}$, and $R_{t}$ denote the number of individuals in the susceptible, infectious, and recovered (removed) compartments, respectively, during the continuous time interval $[t, t + 1)$. The updating equations are thus expressed as:
\begin{align}
\nonumber S_{t+1} &= S_{t} - I_{t}^{*}, \\
I_{t+1} &= I_{t} + I_{t}^{*} - R_{t}^{*}, \label{eq:ODE} \\
\nonumber R_{t+1} &= R_{t} + R_{t}^{*}. 
\end{align} 
Beyond the compartments $S_{t}$, $I_{t}$, and $R_{t}$, the model also includes transition vectors $I_{t}^{*}$ and $R_{t}^{*}$, representing the number of individuals moving into the infectious and recovered compartments during a given time interval. Within a Bayesian framework, the connection between observed data and model parameters is established via probability distributions. Specifically, in SIR models, the transitions between compartments are modeled as binomial random variables, described as follows:
\begin{align}
\nonumber I_{t}^{*} & \sim \text{Bin}(S_{t}, \pi_{t}^{(SI)}), \\
R_{t}^{*} & \sim \text{Bin}(I_{t}, \pi^{(IR)}), \label{eq:SIRmodel} 
\end{align} 
where $\pi_{t}^{(SI)}$ represents the probability of a susceptible individual becoming infected at time $t$, while $\pi^{(IR)}$ indicates the probability of an infected individual recovering at time $t$, the latter being related to the duration of the infectious period. Assuming that contacts occur according to a Poisson process, implying independent interactions at a constant rate, 
the transmission probability $\pi_{t}^{(SI)}$ can be mathematically formulated accordingly:
\begin{equation}
\pi_{t}^{(SI)} = 1 - \exp \left\{ - \beta \frac{I_{t}}{N} \right\}.
\label{eq:transProbSIR} \end{equation} 
The transmission probability equation involves $\beta$, which is the constant transmission rate, and $N$, which denotes the total population size. Here, $\beta$ encompasses both the contact frequency and the likelihood of transmission during contact. In this context, due to the combined effect, these two aspects cannot be distinguished separately. 

Complementarily, the removal probability, represented by $\pi^{(IR)}$, is calculated assuming that the infectious period follows an exponential distribution characterized by the rate parameter $\gamma$. As a result, the removal probability is expressed as:
\begin{equation}
\pi^{(IR)} = 1 - \exp \left\{ - \gamma \right\}.
\label{eq:removProb} \end{equation} 
Although $\gamma$ represents the removal rate, it is often more intuitive to interpret its inverse, $1/\gamma$, which corresponds to the average duration of the infectious period.


\subsection{Behavioral change model} \label{subsec:BC}

The approach proposed by \cite{Ward2023} for the BC model involves incorporating a time-varying transmission mechanism that captures changes in the population's behavior. Although this concept has been relatively underexplored in epidemic modeling literature, it recognizes that the at-risk population modifies its behavior in response to the outbreak's severity, thereby dynamically altering transmission dynamics. In modeling epidemics with the BC framework, this behavioral change is integrated into Eq.~\eqref{eq:transProbSIR} by introducing a component modulated by a time-varying alarm level, denoted as $a_{t}$. The variable $a_{t} \in [0, 1]$ represents the proportionate reduction in transmission due to the population's level of alarm. Specifically, when $a_{t} = 0$, the population is unalarmed, and transmission proceeds at the baseline rate $\beta$. Conversely, when $a_{t} = 1$, the population is fully alarmed, effectively halting transmission entirely. Under these conditions, the traditional transmission probability in Eq.~\eqref{eq:transProbSIR} is reformulated to account for this behavioral response as follows:
\begin{equation}
\pi_{t}^{(SI)} = 1 - \exp \left\{ - \beta (1 - a_{t}) \frac{I_{t}}{N} \right\}.
\label{eq:transProbBC} \end{equation} 
It is worth noting that when $a_{t} = 0$, the BC model no longer allows variation in population behavior, reducing its formulation to that of the traditional SIR model as previously described in Eqs.~\eqref{eq:ODE}--\eqref{eq:removProb}. Furthermore, although the BC model incorporates behavioral changes, it assumes these changes are homogeneous across the entire population, with $a_{t}$ affecting everyone equally.


There are several approaches to defining the alarm function \citep{Ward2023, Ward2024}. At the population level, \cite{Ward2023} suggested defining the alarm function as a function of the smoothed incidence observed over the previous $m$ days. To do so, let $\bar{I}_{[t,t-m]}^{*}$ represent the average of the observed incidences over the past $m$ days, i.e., $\bar{I}_{[t,t-m]}^{*} = \frac{1}{m} \sum_{i = 0}^{m-1} I^{*}_{t-i}$, for $m \leq t$. When $t < m$, we use the moving average of the data up until time $t$. Thus, the alarm function is expressed as $a_{t} = f(\bar{I}_{[t,t-m]}^{*})$. Then, based on the smoothed incidence data observed, the function can be specified in various ways. It is worth noting that although this study focuses on incidence-based alarm functions, it is possible to formulate them based on other epidemic severity metrics, such as prevalence, hospitalizations, or test positivity rates.

A commonly used approach is the one-parameter growth function, also known as the power alarm function \citep{Eksin2019, Ward2023, Ward2024}. This alarm function is defined as: 
\begin{equation}
a_{t} = 1 - \left( 1 - \frac{\bar{I}_{[t,t-m]}^{*}}{N} \right)^{1/k}, \label{eq:power}
\end{equation} 
where $k > 0$ controls the growth rate, with smaller values of $k$ correspond to a more rapid increase in the alarm level. An alternative is the threshold alarm function, which is a two-parameter constant change point function \citep{Ward2023, Ward2024}. This alarm is specified as:
\begin{equation}
a_{t} = \begin{cases} \delta, \quad \text{if} \; \; \frac{\bar{I}_{[t,t-m]}^{*}}{N} > H, \\
    0, \quad \text{otherwise}, \end{cases} \label{eq:thresh}
\end{equation}
where $H$ is the threshold and $\delta \in [0, 1]$ is the infectivity reduction factor. Fundamentally, this function remains equal to zero until a specified threshold is exceeded, at which point it assumes a value of $\delta$. Another alarm function approached in this study is a two-parameters Hill-type alarm \citep{Kassa2011, Ward2024}, which is defined as:
\begin{equation}
a_{t} = \frac{ \left( \frac{\bar{I}_{[t,t-m]}^{*}}{N} \right)^{\nu}}{\delta^{\nu} + \left( \frac{\bar{I}_{[t,t-m]}^{*}}{N} \right)^{\nu}}. \label{eq:hill}
\end{equation} 
In this two-parameters function, $0 < \delta < 1$ represents the proportion of the population that must be infectious to achieve half of the maximum BC effect, while $\nu > 0$ is the Hill coefficient. As $\nu$ increases, the Hill function produces a more gradual onset of the BC effect compared to the other two smooth alarm functions previously discussed. As $\nu \rightarrow \infty$, the Hill alarm function approaches the behavior of a threshold alarm. 


\section{Model formulation} \label{sec:model}

The aim here is to develop a model that captures heterogeneity in population behavior. Thus, we propose a Bayesian mixture model for behavioral dynamics. Specifically, our approach distinguishes between two subgroups within the population: one group that maintains consistent behavior throughout the epidemic and another group that changes behavior in response to the epidemic severity. To do so, let $\omega$ denote the proportion of the total susceptible population that remains in its natural, unalarmed state when confronted with an epidemic, that is, the risk-neutral group. Conversely, $1 - \omega$ represents the proportion of individuals who, in response to the epidemic, alter their behavior and become alarmed, reflecting a risk-averse attitude. Taking this into account, we reformulate the transmission probability equation to incorporate both groups accordingly:
\begin{equation}
\pi_{t}^{(SI)} = 1 - \exp \left\{ - \beta \big[ \omega \;  + (1 - \omega) \; (1 - a_{t}) \big] \frac{I_{t}}{N} \right\}.
\label{eq:transProbMIX} \end{equation}  
As in the SIR and BC models, $\beta \in [0, \infty)$ denotes the transmission rate, while $a_{t} \in [0, 1]$ represents the alarm function. Eq.~\eqref{eq:transProbMIX} effectively integrates Eq.~\eqref{eq:transProbSIR} from the traditional SIR model with Eq.~\eqref{eq:transProbBC} from the BC model, thereby capturing potential heterogeneity in population behavior. When $\omega = 1$, the model assumes that the entire population remains in its unalarmed state, reducing to the traditional SIR model (Eq.~\eqref{eq:transProbSIR}). Conversely, when $\omega = 0$, the entire population adjusts its behavior in response to the epidemic, making the model equivalent to the standard BC model (Eq.~\eqref{eq:transProbBC}). More importantly, for values of $0 < \omega < 1$, the model captures a mixture of two distinct behavioral patterns within the population.

\subsection{Prior specifications} \label{subsec:prior}

Following the Bayesian paradigm, prior probability distributions must be assigned to all model parameters. For SIR models, it is well established that transitions between compartments follow a binomial distribution (see Section~\ref{subsec:SIR}). Consequently, as shown in Eqs.~\eqref{eq:removProb} and \eqref{eq:transProbMIX}, prior distributions must be specified for the parameters $\beta$ and $\gamma$ to fully define the Bayesian model. Gamma distributions are commonly used as priors for both $\beta$ and $\gamma$ because their support lies on the positive real line. When prior knowledge is available regarding the infectious period, informative prior distributions may be specified for $\gamma$. Additionally, prior information about the basic reproduction number, defined as $R_0 = \beta/\gamma$, can guide the choice of hyperparameters for the prior distribution of $\beta$. In the absence of such information, weakly informative priors are typically assigned to both parameters. In this study, Ga$(a, b)$ represents a gamma distribution with mean $a/b$.

In the case of BC models, choosing prior distributions for the parameters of the alarm functions can be more challenging since identifying how people will react to an outbreak and how this reaction affects transmission rates {\it a priori} is difficult. In such situations, vague or weakly informative priors are generally preferred. For the parameter $k$ in the power alarm function (Eq.~\eqref{eq:power}) and $\nu$ in the Hill-type alarm function (Eq.~\eqref{eq:hill}), a gamma distribution is appropriate, as these parameters must take positive values. For the parameter $\delta$ in the threshold and Hill-type alarm functions (Eqs.~\eqref{eq:thresh} and \eqref{eq:hill}, respectively), a flat distribution on the interval [0, 1] is suitable. Ultimately, for $H$ in the threshold alarm function (Eq.~\eqref{eq:thresh}), a uniform prior defined over the observed range of incidence can be used, as we expect the change point to occur within this interval. Note that this strategy could generate some inconvenience if this setting is too informative.

Finally, when fitting a mixture model, in addition to specifying priors for the previously discussed parameters, a distribution must also be assigned to the weight parameter $\omega$. Given that prior knowledge about population behavior is often available, it is both possible and appropriate to set an informative prior. The beta distribution, denoted Be$(a, b)$, is well suited for $\omega$ since it models proportions constrained to the interval [0, 1]. In the absence of such prior information, a beta specification equivalent to a standard uniform distribution on the interval [0, 1] can be used.

\subsection{Posterior inference} \label{subsec:infe}

Once the model and priors are properly defined, we next describe the procedure for estimating the parameters of the proposed model given the observed data. Following Bayes’ theorem, the joint posterior distribution of the parameters is proportional to the product of the likelihood and the prior distributions, which, for the proposed mixture behavior model, is given by:
\begin{align}  
p({\bm \Omega} \mid \bm{I}^{*}, \bm{R}^{*}) \propto & \prod_{t = 1}^{T} \binom{S_{t}}{I_{t}^{*}} \left(\pi_{t}^{(SI)} \right)^{I_{t}^{*}} \left( 1 - \pi_{t}^{(SI)} \right)^{S_{t} - I_{t}^{*}} \; \binom{I_{t}}{R_{t}^{*}} (\pi^{(IR)})^{R_{t}^{*}} (1 - \pi^{(IR)})^{I_{t} - R_{t}^{*}} \times p({\bm \Omega}),
\label{eq:posterior} \end{align} 
where $\pi_{t}^{(SI)}$ and $\pi^{(IR)}$ are defined in Eqs.~\eqref{eq:transProbMIX} and \eqref{eq:removProb}, respectively. The parameter set is denoted by ${\bm \Omega}$ and includes the transmission and removal rates ($\beta$ and $\gamma$), the weight parameter ($\omega$), and the alarm parameters according to the specified form of the alarm function ($k$ for the power function; $\delta$ and $H$ for the threshold function; and $\delta$ and $\nu$ for the Hill-type function). Finally, $p({\bm \Omega})$ denotes the joint prior distribution, as discussed in Section~\ref{subsec:prior}.

All inference in this study is performed using MCMC methods via the R package {\tt nimble} \citep{Valpine2017}. In addition to estimating parameters, we are interested in conducting posterior predictive inference. By simulating new data values, one for each of a number of posterior draws, we obtain an approximation of the posterior predictive distribution, which reflects uncertainty in future observations by averaging over all possible parameter values weighted by their posterior probabilities. This distribution enables model validation through comparison of predicted and observed data and quantifies predictive uncertainty by incorporating variability in the model parameters. Formally, the posterior predictive distribution for the mixture behavior model is given by:
\begin{align} 
p( \tilde{\bm{I}}^{*}, \tilde{\bm{R}}^{*} \mid \bm{I}^{*}, \bm{R}^{*}) &= \int_{\bm \Omega} L( \tilde{\bm{I}}^{*}, \tilde{\bm{R}}^{*} \mid {\bm \Omega}) \; 
p({\bm \Omega} \mid \bm{I}^{*}, \bm{R}^{*}) \; d{\bm \Omega},
\label{eq:predictive} \end{align} 
where $L( \tilde{\bm{I}}^{*}, \tilde{\bm{R}}^{*} \mid {\bm \Omega})$ is the likelihood of the new data, and $p({\bm \Omega} \mid \bm{I}^{*}, \bm{R}^{*})$ is the posterior distribution of the parameters given the observed data, as presented in Eq.~\eqref{eq:posterior}. Although computing this posterior predictive distribution analytically is not feasible, MCMC sampling provides an effective approximation. The resulting simulated datasets are then compared with the observed data as part of posterior predictive checking, offering a practical means to assess whether the model assumptions are consistent with empirical patterns.

Note that the key advantage of adopting a Bayesian framework is that uncertainty quantification is naturally and coherently incorporated into all stages of the analysis. By treating unknown parameters as random variables with well-defined probability distributions, the Bayesian approach propagates uncertainty from the estimation stage through to forecasting in a principled manner. As a result, posterior distributions provide a complete characterization of the uncertainty of the estimation procedure, while posterior predictive distributions yield predictions that explicitly reflect both parameter and observational uncertainty. This unified treatment leads to more transparent inference and more reliable predictive performance, particularly in settings where uncertainty plays a central role in decision-making.

\section{Simulation studies} \label{sec:simulations}

This section aims to gauge the properties of the proposed model in different aspects. First, we seek to determine whether the behavioral change mechanism can be recovered by estimating the alarm function through the mixture model. Additionally, we compare the proposed model with the traditional SIR (which does not incorporate behavioral change) and homogeneous BC models, assessing their posterior predictive performance, model fit, and accuracy of future predictions. These objectives were addressed by simulating epidemics under five different conditions: (i) without behavioral change (i.e., $\omega = 1$); (ii) with behavioral change driven by the power alarm function (i.e., $\omega = 0$); and with heterogeneity in population behavior characterized by (iii) 30\% risk-neutral and 70\% risk-averse individuals (i.e., $\omega = 0.3$), (iv) 50\% risk-neutral and 50\% risk-averse individuals (i.e., $\omega = 0.5$), and (v) 70\% risk-neutral and 30\% risk-averse individuals (i.e., $\omega = 0.7$). For scenarios (iii)–(v), the risk-neutral proportion represents a segment of the population following the traditional SIR model ($\omega$), while the risk-averse group has transmission rates dictated by the behavioral change model under the power alarm function ($1 - \omega$). For each of these five scenarios, we simulated 100 epidemic outbreaks using the initial conditions: total population $N = 1,000,000$, initial susceptible number $S_0 = 999,995$, and initial infective number $I_0 = 5$. The parameters were set to $\gamma = 0.143$, $\beta = 0.429$, $k = 0.1$, with $\omega$ taking values of $1$, $0$, $0.3$, $0.5$, and $0.7$ according to the scenario. These settings produce epidemics with an average infectious period of approximately 7 days and a basic reproduction number around 3. For epidemics generated by both the BC and mixture models, a 14-day average incidence was used to inform the alarm function.

\begin{figure}[t] \centering 
    \includegraphics[width=0.4\textwidth]{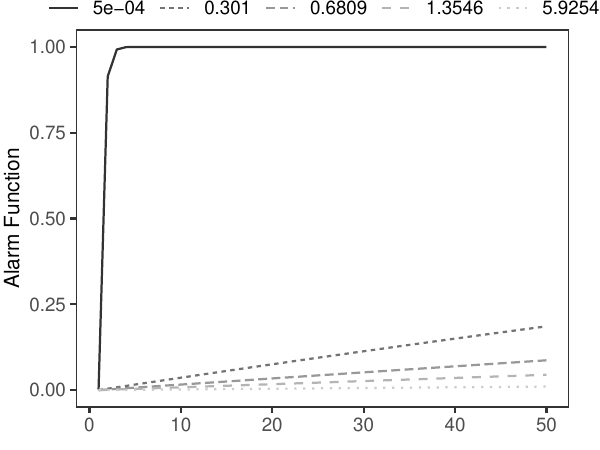}
    \caption{Power alarm function evaluated on the quantiles corresponding to the probabilities (0.00, 0.25, 0.50, 0.75, 1.00) of the sample of $k$ generated from the prior distribution $k \sim \text{Ga}(1, 1)$.}
    \label{fig:proiorAlarm}
\end{figure}

To examine the properties of the proposed model, we fitted three different models to each simulated epidemic: a traditional SIR model, a standard BC model using the power alarm function, and the proposed mixture model that combines both. Additional simulation studies based on threshold and Hill-type alarm functions can be found in the Appendix. Priors were specified following the recommendations in Section~\ref{subsec:prior}. The prior for $\gamma$ was set to $\text{Ga}(20.45, 143)$, corresponding to a mean infectious period of seven days and implying an 80\% prior probability that the mean lies between 5 and 9 days. For $\beta$, a $\text{Ga}(1.84, 4.29)$ prior was assumed, which yields an 80\% prior probability that the mean transmission rate falls between 0.11 and 0.85. Under these priors for $\gamma$ and $\beta$, the basic reproduction number lies in the range $[0.73, 6.37]$ with 80\% prior probability. Weakly informative priors were assigned to the alarm parameter $k$ and the weight $\omega$, specifically $\text{Ga}(1, 1)$ and $\text{Be}(1, 1)$, respectively. The prior assumed for $k$ allows the alarm function to take values across the entire interval. To illustrate this, we generated a sample from the $\text{Ga}(1, 1)$ distribution, computed the sample quantiles corresponding to the probabilities (0.00, 0.25, 0.50, 0.75, 1.00), and evaluated the alarm function at these values (Figure~\ref{fig:proiorAlarm}). Regarding the weights, the $\text{Be}(1, 1)$ prior assigns equal probability to all possible values of $\omega$. 

All models were run by extracting 10,000 samples from a total of 1,000,000 iterations, with the initial 80\% of samples discarded as burn-in and a thinning factor of 20 applied to reduce autocorrelation. Convergence was evaluated visually using graphical diagnostics and numerically by assessing the effective sample size \citep[ESS;][Chapter 11]{Gelman2013}. Convergence was judged to have failed for two SIR replicates and one mixture replicate (iv) because at least one parameter had an ESS below 200. Therefore, these replicates were excluded from the analysis. The remaining replicates in this scenario exhibited a minimum ESS greater than 1,000, while the minimum ESS observed for the other scenarios exceeded 3,000. In this section, we focus on exploring specific characteristics of the proposed model, but additional results are provided in the Appendix.

\subsection{Alarm function and parameter estimation}

\begin{figure}[t] \centering 
    \includegraphics[width=\textwidth]{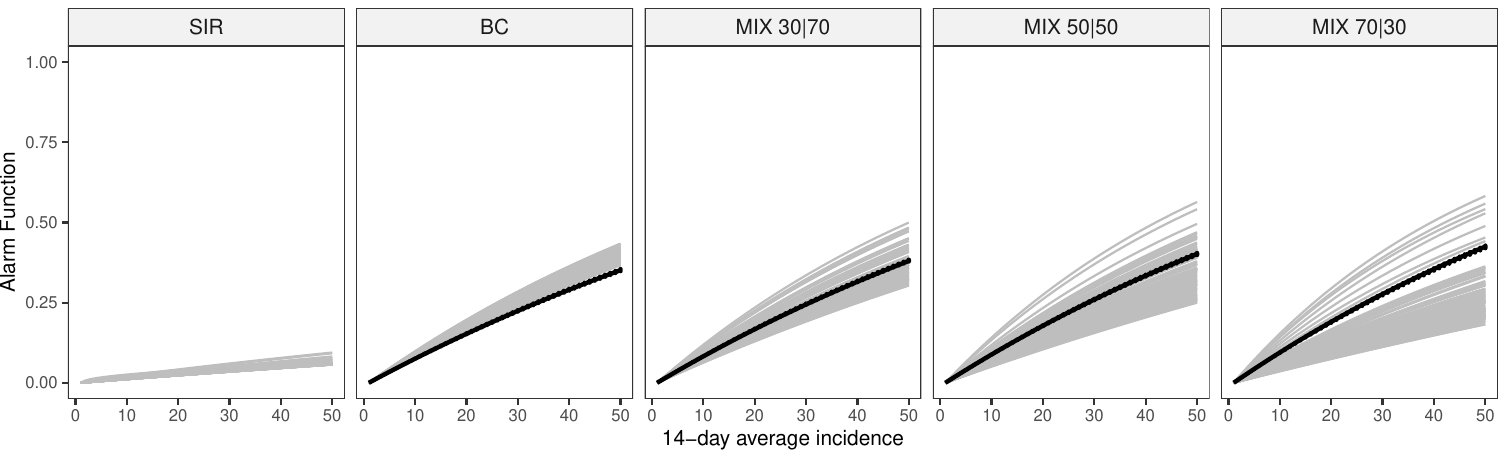}
    \caption{True alarm function (black lines) and posterior mean estimates of alarm functions (gray lines) estimated by the mixture model. Results for 89 simulated epidemics using the traditional SIR model (SIR), behavioral change model (BC), and mixture models with different proportions of risk-neutral populations: 30\%, 50\%, and 70\% (MIX 30$|$70, MIX 50$|$50, and MIX 70$|$30, respectively).}
    \label{fig:SimAlarm}
\end{figure}

To assess the accuracy of the estimated alarm function across different data-generating scenarios, we compare the posterior mean estimates of the alarm function with the true alarm function (see Figure~\ref{fig:SimAlarm}). The curve depicts the alarm as a function of the 14-day average incidence, ranging from zero to the maximum observed during the epidemic, which varies across simulations. When the mixture model is fitted to data generated from the traditional SIR model, the alarm function provides little information because the weight $(1 - \omega)$ multiplying the alarm function in Eq.~\eqref{eq:transProbMIX} is zero. In other words, under the true model, no individuals follow the BC model, rendering the alarm function irrelevant. A similar pattern appears in data generated from the mixture model when the majority of the population (70\%) is risk-neutral, which explains the substantial variability observed in the estimated function. Conversely, in scenarios where the true model incorporates the alarm function for at least 50\% of the population, the mixture model estimates the alarm function more accurately. This precision increases as the proportion of the risk-averse population grows. 

We also assess the models' ability to estimate other parameters. Detailed results are provided in the Appendix. The traditional SIR model accurately recovers parameters only when data was simulated directly from it. In other scenarios, this model tends to result in underestimates of $\gamma$ and overestimates of $\beta$. A similar pattern is observed with the standard BC model. When fitted to data generated from itself, the fitted model successfully recovers the true parameters. However, when applied to data from other models, the BC model overestimates the alarm parameter $k$, possibly to account for a risk-neutral population. In contrast, the mixture model performs well across all scenarios. To evaluate its accuracy, 95\% credible intervals were computed, and the coverage rate was estimated (see Appendix Table\ref{SM:tab:PowerSummary}). When data were generated from the traditional SIR model, the model estimates $\omega \in [0.47, 1.00]$, with a coverage rate of 0.74, meaning that the true value is within the 95\% credible interval for 74\% of the replicates. For data generated from the standard BC model, $\omega \in [0.00, 0.24]$, with a coverage rate of 0.73. It is worth noting that the coverage rates in these cases are relatively low, as the true values of $\omega$ lie near the boundaries of the distribution, which hinders precise estimation. When data were generated from the mixture model itself, the estimates of $\omega$ are approximately [0.04, 0.45], [0.06, 0.61], and [0.07, 0.76], with coverage rates of 0.99, 0.94, and 0.84, respectively, corresponding to scenarios in which 30\%, 50\%, and 70\% of the population is risk-neutral. It is worth noting that we used a flat prior distribution for the weight parameter, $\omega \sim \text{Be}(1,1)$, which does not provide the model with guidance in estimating this parameter. Using more informative priors for the weight may well be feasible in practice and would likely improve the model’s coverage for $\omega$ and other parameters. Nevertheless, the results demonstrate that the mixture model is sufficiently flexible to accurately fit data across a variety of scenarios.

\subsection{Posterior prediction}

\begin{figure}[t] \centering 
    \includegraphics[width=\textwidth]{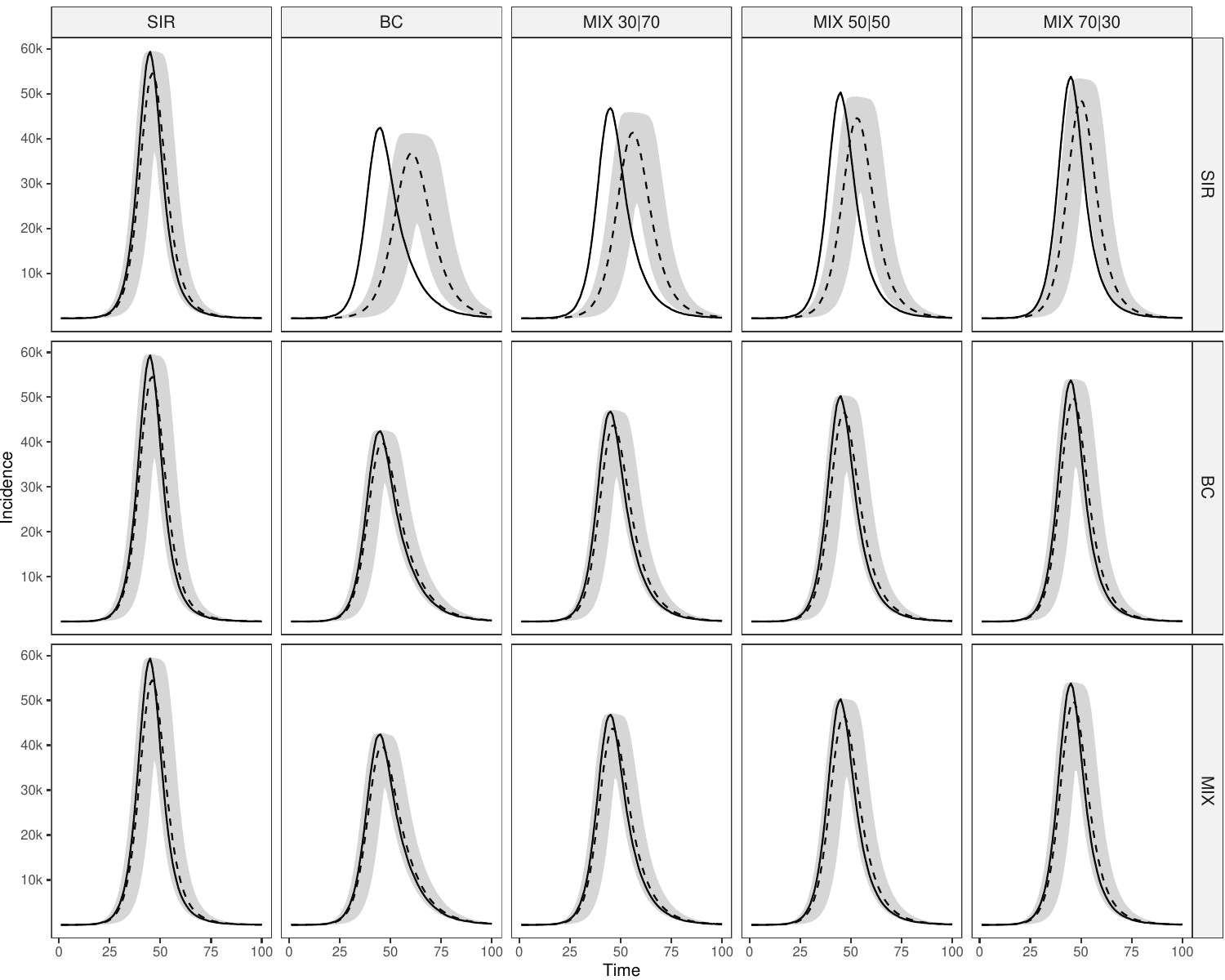}
    \caption{True incidence curve (solid line), posterior predictive mean (dashed line), and 95\% credible intervals for a randomly selected simulation from different mechanisms. Rows correspond to fitted models, and columns correspond to simulation models.}
    \label{fig:SimEst}
\end{figure}

The ability of the fitted models to estimate the epidemic incidence curve was assessed by comparing it with realizations from the posterior predictive distribution. Specifically, the epidemic trajectory was evaluated across the 10,000 posterior samples as an approximation of the posterior predictive distribution presented in Eq.~\ref{eq:predictive}. The average incidence curve derived from these samples was plotted along with its 95\% posterior credible intervals. The results are consistent across simulations. To illustrate this, the posterior predictive distribution for one representative simulation is shown in Figure~\ref{fig:SimEst}. Overall, the SIR model performs well when the data were generated from that same model. In the other scenarios, the posterior predictive incidence curves exhibit a delay in reaching their peaks, with this delay decreasing as the proportion of the risk-neutral population increases. Differences between the predictions obtained from the BC and mixture models are minimal, as both models adapted effectively to various incidence curve shapes. 

Model performance is assessed using the widely applicable information criterion \cite[WAIC;][]{Watanabe2010}, where lower values indicate a better fit to the data. Boxplots of the WAIC values obtained in each scenario are provided in Appendix Figure~\ref{SM:fig:WAICPower}. Overall, these findings are consistent with those observed in the posterior predictive plots. The SIR model performs similarly to the others only when the data are generated from itself, in all other scenarios, its performance is worse. BC and mixture models show comparable performance across all scenarios.

\subsection{Forecasting}

\begin{figure}[!t] \centering 
    \includegraphics[width=\textwidth]{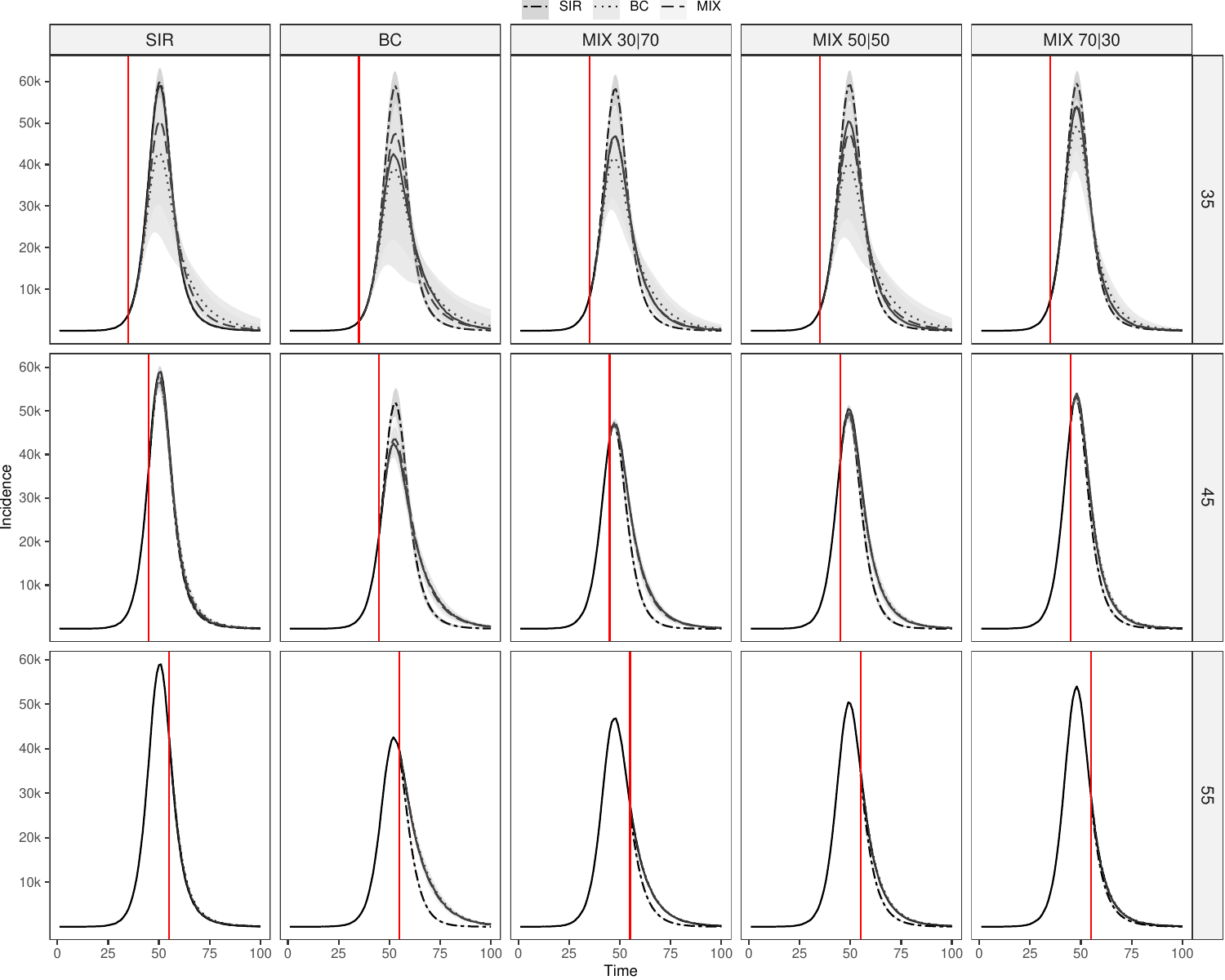}
    \caption{Posterior predictive mean (dashed, dotted, and long-dashed lines for SIR, BC, and mixture models, respectively) and 95\% credible intervals for forecasts of future incidence compared to the true values (solid line) for a randomly selected simulation. Results were generated considering different cutoffs at days 35, 45, and 55. BC and MIX curves are overlapping.}
    \label{fig:SimFor}
\end{figure}

The predictive performance of the fitted models is evaluated using the posterior distribution of the model parameters inferred from the early phase of the epidemic. To provide a more comprehensive assessment, we selected several cutoff points to examine how well the models predict the epidemic trajectory before, during, and after its peak. In particular, we used cutoffs at days 35, 45, and 55, fitted each model to data available up to the corresponding day, and then used the remaining observations to evaluate forecasting performance. For each cutoff, we simulated the future trajectory of the epidemic starting from the model state on that day, drawing on 10,000 posterior samples of the parameters estimated from data up to the respective cutoff (i.e., $t = 1, \ldots, \text{cutoff}$). 
Based on these draws, we then plotted the mean and the 95\% credible intervals for the forecasts of future incidence. The simulations produced consistent results across different replicates. For clarity, Figure~\ref{fig:SimFor} presents the posterior predictive distribution from a single, randomly selected simulation. 

The SIR model produces accurate forecasts when applied to data generated from the same model, regardless of the chosen cutoff point. However, its performance varies when forecasting data generated by alternative approaches. When the cutoff was set early in the epidemic (day 35), the SIR model overestimates the epidemic peak, leading to a steeper decline in the predicted epidemic curve compared to the true trajectory. With the cutoff closer to the peak (day 45), the model provides a more accurate estimate of the peak itself, yet it still predicted a faster decline than observed. This tendency persisted even when the cutoff was set after the peak (day 55). In contrast, the BC model generates wider credible intervals but generally outperformed the SIR model in forecasting accuracy. The only scenario in which the BC model fails to capture the true epidemic trajectory is when the data were simulated using the SIR model with a cutoff at day 35. Forecast accuracy improved for all models as the cutoff point moved further along the epidemic timeline. Lastly, we assess the performance of the mixture model. This model consistently predicts the true epidemic trajectory across all simulation scenarios and cutoff points. Moreover, it produced more accurate credible intervals than the BC model.

\section{Application} \label{sec:application}

In this section, we demonstrate the practical utility of the proposed method by applying it to data from the COVID-19 pandemic, including the New York City and Montreal COVID‑19 data described in \cite{Ward2023, Ward2025}. 

\subsection{COVID-19}

To illustrate the model’s capacity to capture heterogeneous transmission dynamics across distinct urban settings, we analyze reported COVID‑19 case data from New York City (New York, United States) and Montreal (Quebec, Canada). For New York City, we examine two epidemic waves occurring between March 2020 and May 2021, and for Montreal, we analyze two waves spanning March 2020 to March 2021. These cities offer a useful comparative framework because their epidemic trajectories differ markedly over time. In both locations, the first wave was relatively short (108 days in New York City and 127 days in Montreal) and exhibited a single peak. The second wave was longer and more complex, lasting 227 days in New York City and 210 days in Montreal, presenting multiple peaks. Population sizes were set to $N = 8,804,190$ for New York City, based on the 2020 U.S. Census, and $N = 1,762,949$ for Montreal, as reported in the 2021 Canadian Census of Population.

To compare the behavior of the different model specifications, we fitted the traditional SIR model, the standard BC model with power (Eq.~\eqref{eq:power}), threshold (Eq.~\eqref{eq:thresh}), and Hill‑type (Eq.~\eqref{eq:hill}) alarm functions, and the corresponding mixture models to each COVID‑19 wave, all formulated as previously presented in Sections~\ref{sec:back} and \ref{sec:model}. Vague priors were assigned to the alarm-function parameters ($k \sim \text{Ga}(0.1, 0.1)$, $\delta \sim \text{Be}(1,1)$, $H \sim \text{U}(\min(I_{t}),\max(I_{t}))$, and $\nu \sim \text{Ga}(1, 0.1)$). 
The prior for $\gamma$  was chosen to reflect a mean infectious period of approximately three days, with 80\% prior probability that the mean lies between 2 and 4 days, which is consistent with the typical interval between becoming contagious, testing positive, and subsequently isolating. For $\beta$, we specified a prior that yields an 80\% prior probability that the mean transmission rate falls between 0.1 and 2.3. Under these priors for $\gamma$ and $\beta$, the basic reproduction number has an 80\% prior probability of lying within the range $[0.3, 7.5]$. Weakly informative priors were assigned to the mixture weights after performing a brief sensitivity analysis comparing priors centered at 0.3, 0.5, and 0.7, which showed minimal differences in results. Therefore, we adopted a Be$(5,5)$ prior for $\omega$, implying an 80\% prior probability that the proportion of the population remaining in its natural, unalarmed state lies between 0.3 and 0.7, thereby avoiding extreme values. Following \cite{Ward2023}, the 30-day average incidence was used to inform the alarm functions. All models were run by extracting 10,000 samples from a total of 1,000,000 iterations with the initial 80\% of samples discarded as burn-in. A thinning factor of 20 was applied to reduce the sample. Three independent chains were considered for the MCMC in order to provide more reliable results.

Prior to analyzing the results, MCMC convergence was evaluated through visual inspection of standard graphical diagnostics and quantified numerically using the ESS and the Gelman-Rubin statistic \citep[$\hat{R}$;][]{Gelman1992}. Detailed convergence diagnostics are provided in the Appendix. Despite the high number of iterations, the BC model featuring a Hill-type alarm function sometimes failed to achieve satisfactory convergence. Specifically, this occurred during Wave 1 for New York City and Wave 2 for Montreal. As the most flexible functional form explored in this study, the Hill-type alarm function likely attempts to compensate for unobserved population heterogeneity, leading to identifiability issues and poor convergence. Evidence for this is given by the fact that when the same Hill-type function is implemented within a mixture model framework, convergence is successfully achieved. 

To determine the optimal model specifications for the analyzed COVID‑19 waves, we compare performance using the WAIC, as summarized in Table~\ref{tab:WAIC}. The standard SIR model consistently yields the highest WAIC across all waves and both cities, underscoring the need to incorporate behavioral‑change mechanisms when modeling empirical epidemic data. When comparing the standard and mixture BC models, the mixture formulations generally exhibit better or equivalent performance relative to their standard counterparts. This indicates that the mixture model is appropriate. Although more flexible and complex, it achieves lower WAIC values. The mixture BC model featuring a Hill‑type alarm function achieves the lowest WAIC for the first wave in both locations. For the second wave, the mixture and standard BC models featuring a threshold alarm function yield equivalent WAIC values.
\begin{table}[!t] \centering
\caption{WAIC values for all converged models used in the COVID-19 analysis for New York City and Montreal. Lower WAIC values indicate better fit. \label{tab:WAIC}}
\begin{tabular}{ccccc}
\toprule
\multirow{2}{*}{Model fitted} & \multicolumn{2}{c}{New York City} & \multicolumn{2}{c}{Montreal} \\
\cmidrule{2-5}
  & Wave 1 & Wave 2 & Wave 1 & Wave 2 \\
\midrule
SIR & 2265.5 & 3310.5 & 966.6 & 1859.5 \\
BC-Power & 1753.4 & 3296.6 & 933.0 & 1844.0 \\
MIX-Power & 1246.0 & 3295.3 & 894.9 & 1844.3 \\
BC-Threshold & 1466.5 & \textbf{3227.7} & 918.5 & \textbf{1836.2} \\
MIX-Threshold & 1465.8 & \textbf{3226.4} & 918.1 & \textbf{1836.2} \\
BC-Hill & -- & 3249.1 & 905.1 & -- \\
MIX-Hill & \textbf{1189.6} & 3246.4 & \textbf{892.0} & 1852.8 \\
\bottomrule 
\end{tabular} \end{table}

\begin{figure}[!t] \centering 
    \includegraphics[width=\textwidth]{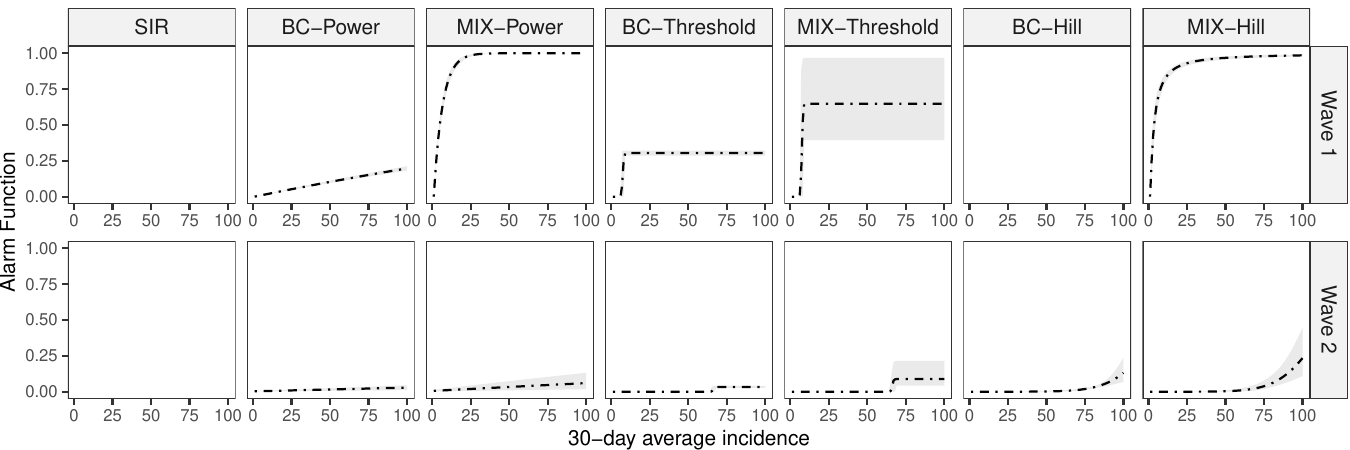} \\
    \includegraphics[width=\textwidth]{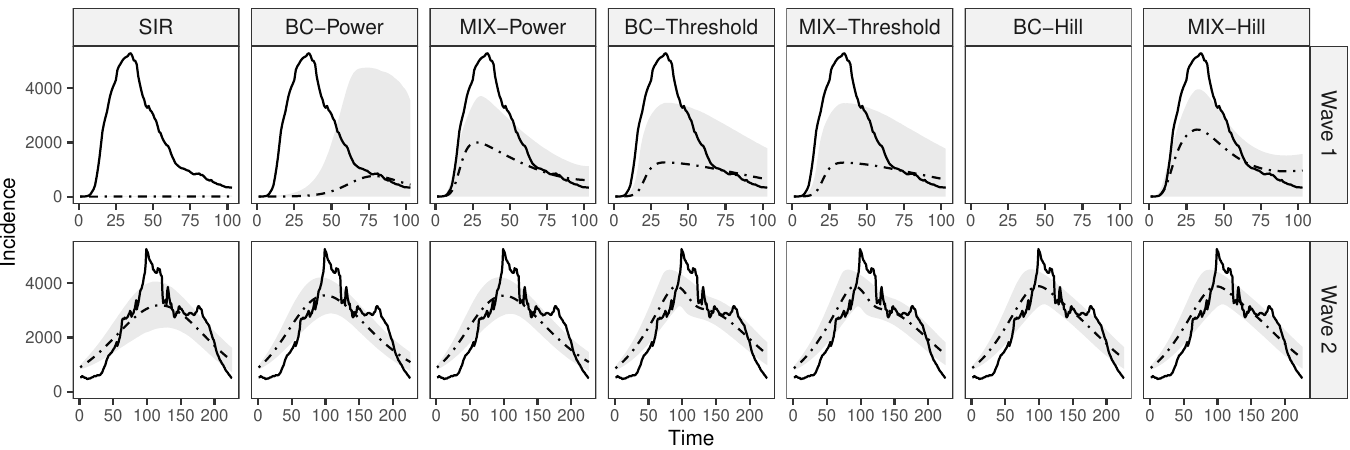}
    \caption{Posterior means and 95\% credible intervals for all estimated alarm functions and posterior predictive distribution. All results from each wave of the New York City COVID-19 epidemic.} \label{fig:NYAlarm}
\end{figure}

\begin{figure}[!t] \centering 
    \includegraphics[width=\textwidth]{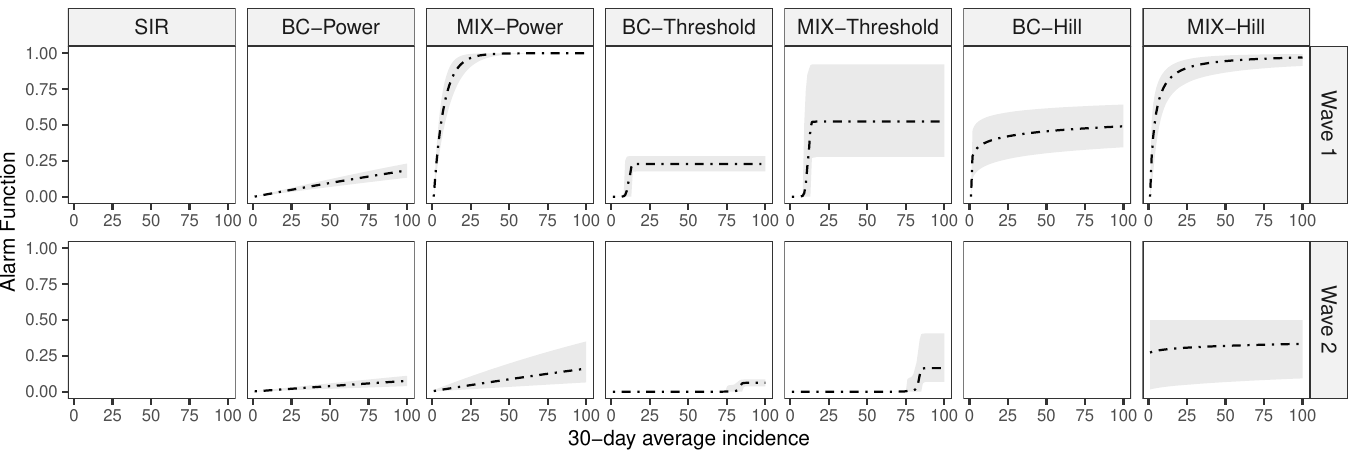} \\
    \includegraphics[width=\textwidth]{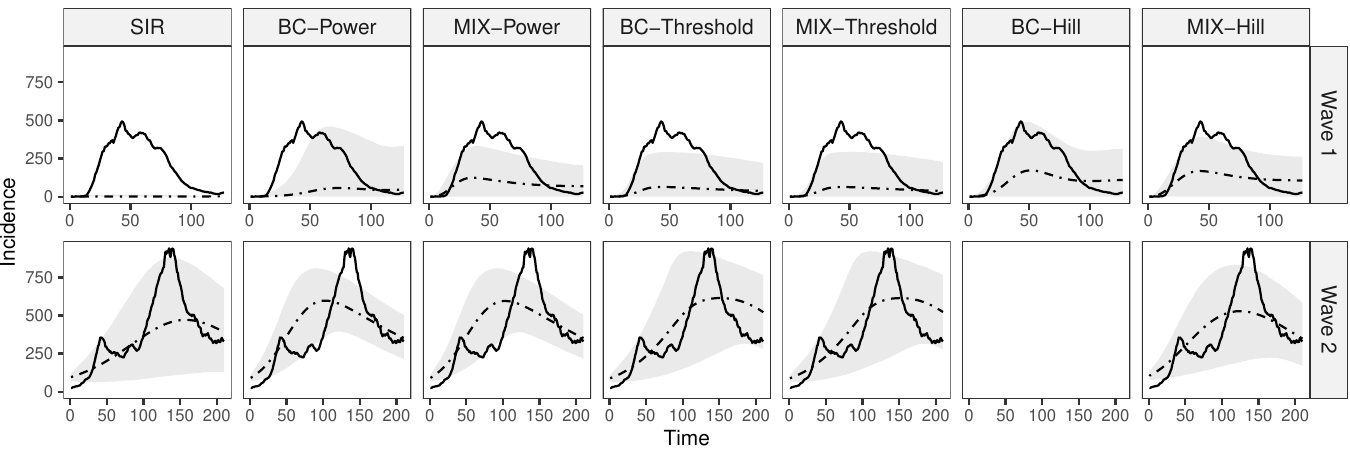}
    \caption{Posterior means and 95\% credible intervals for all estimated alarm functions and posterior predictive distribution. All results from each wave of the Montreal COVID-19 epidemic.} \label{fig:MOAlarm}
\end{figure}

Figures~\ref{fig:NYAlarm} and \ref{fig:MOAlarm} illustrate the estimated alarm functions for the BC and mixture models, derived from the 30-day moving average incidence across all waves and locations. A comparative analysis of these functions reveals distinct behavioral patterns that vary by both the model framework and the specific epidemic wave. During the initial wave, the mixture model generates substantially higher alarm levels than the BC model in both cities. However, this discrepancy is markedly less pronounced during the second wave. Evaluating the alarm functions across successive waves provides critical insights into the temporal evolution of population-level responses. In the first wave in both cities, alarm levels exhibit a sharp increase despite relatively low observed incidence, suggesting high initial sensitivity to the emerging threat. Conversely, at the onset of the second wave, during which many public health restrictions remained in place, both the BC and mixture models estimate only modest increases in alarm, which are triggered only at significantly higher levels of observed incidence.

The final evaluation of the proposed models concerns their capacity to reconstruct the epidemic trajectory, which is assessed using the posterior distributions of the model parameters. Consistent with the simulation study (Section~\ref{sec:simulations}), the epidemic curve is characterized using 10,000 posterior samples to approximate the posterior predictive distribution defined in Eq.~\eqref{eq:predictive}. The resulting mean incidence curves, along with their associated 95\% credible intervals, are presented in Figures~\ref{fig:NYAlarm} and \ref{fig:MOAlarm} for New York City and Montreal, respectively. When evaluating the estimates for the first wave, the traditional SIR model failed to capture the observed dynamics. This failure underscores the necessity of incorporating behavioral change mechanisms into empirical epidemic modeling, a requirement corroborated by the previously discussed WAIC values. In comparing the predictive performance of the BC and mixture models, the most significant discrepancy occurs when employing the power alarm function, where the mixture model demonstrates superior fit. In contrast, this difference became nearly imperceptible when utilizing the threshold or Hill-type alarm functions. Estimates for the second wave exhibit greater consistency across all model specifications, with each framework producing similar structural trends. The primary variations between models in this period are related to the magnitude of posterior uncertainty rather than the trajectory itself. It is worth noting that the alarm functions estimated for the second wave remained consistently low, suggesting that population behavior reaches a state of relative stasis or exhibits minimal reactivity to incidence fluctuations at this stage of the pandemic.

\section{Discussion} \label{sec:discussion}

Methodologically, this paper introduces a novel framework that bridges the gap between classical compartmental models and behavioral change theories by explicitly incorporating behavioral heterogeneity. By partitioning a population into risk-neutral and risk-averse groups using a weight parameter, the model moves beyond the traditional ``behavioral monolith'' assumption, allowing for a more realistic representation of how fragmented social groups respond to a health crisis. This is particularly vital for avoiding biased parameter estimation and ``stretched'' epidemic curves that often plague models assuming population homogeneity. Practically, the model provides public health officials with a more nuanced computational toolkit to predict outbreak trajectories in socially complex environments, ensuring that intervention strategies are informed by behavioral realism rather than aggregate averages.

The results of our simulation studies demonstrate that the proposed Bayesian mixture framework significantly outperforms traditional SIR and standard BC models across several critical metrics. Most notably, the mixture model achieved high-fidelity recovery of true parameters across all five simulated scenarios, ranging from populations with no behavioral adjustment ($\omega=1$) to those with varying proportions of risk-averse individuals. While the traditional SIR model frequently overestimated epidemic peaks and forecasted artificially rapid declines, the mixture approach consistently tracked the true trajectory throughout the epidemic lifecycle. Furthermore, the mixture model provided superior predictive accuracy and more reliable credible intervals compared to standard BC models, which often struggled to calibrate to heterogeneous data. These findings indicate that by explicitly accounting for internal population variance, our model avoids the common pitfalls of biased parameter estimation and inaccurate forecasting, maintaining robust performance even in complex risk-profile scenarios.

The application of the proposed framework to COVID‑19 datasets from New York City and Montreal underscores its robust performance and its utility as a diagnostic tool for deciphering complex social responses to health crises. Beyond theoretical validation, the mixture model demonstrates flexibility in capturing distinct urban transmission dynamics. In comparative analyzes, it consistently outperforms the traditional SIR model and achieves performance that is equivalent to or better than standard BC specifications in terms of predictive fit, as reflected by lower WAIC values. Our findings further reveal a critical temporal shift in population sensitivity: while alarm levels were highly responsive at the pandemic’s onset, they became increasingly resistant to change during subsequent waves, a phenomenon often described as ``pandemic fatigue''. By quantifying the weight parameter ($\omega$), the model provides a rigorous mechanism for identifying the proportion of the population that remains ``unalarmed'' by rising incidence. Consequently, this framework offers public health officials a realistic lens through which to interpret multilayered datasets, assess the effectiveness of past interventions, and refine future strategies in sociologically diverse metropolitan environments.

Building upon the foundational framework established in this study, several promising avenues for future research emerge to further refine the modeling of behavioral heterogeneity. First, while the current approach partitions the population into two distinct groups, future iterations could incorporate a continuous spectrum or a larger set of behavioral classes to better reflect the granular social fragmentation observed in empirical settings. Furthermore, integrating these mixture dynamics into spatial models or individual-level modeling \citep{Deardon2010, Rahul2024, Ward2024} could enhance the precision of localized outbreak predictions across geographically diverse regions, accounting for the interplay between movement patterns and risk tolerance. Finally, a critical extension involves the development of dynamic weight parameters. Rather than assuming a constant distribution of behavioral types, investigating how the mixing proportion $\omega$ evolves over time could capture shifting social sentiments, such as the transition from risk-aversion to risk-neutrality driven by ``pandemic fatigue''. Such advancements would provide a more robust and adaptive toolkit for public health officials to monitor and respond to the long-term trajectories of infectious diseases.

\section*{Acknowledgements}

This research was enabled in part by support provided by the Research Computing Services group at the University of Calgary. A. M. Schmidt and R. Deardon acknowledge financial support from the Natural Sciences and Engineering Research Council (NSERC) of Canada Discovery Grants program (Schmidt—RGPIN/2024-04312, Deardon—RGPIN/03292-2022).

\bibliographystyle{abbrvnat}
\bibliography{reference}

\newpage
\appendix
\section*{Appendix}

This appendix is organized as follows. Section~\ref{SM:sim} presents additional simulation studies conducted to further examine the properties of the model. Section~\ref{SM:app} provides complementary results that support the application of the model to COVID‑19 datasets from New York and Montreal. 

Data and R code: https://github.com/jlpavani/Heterogeneous-behavioral-change-modeling includes all opensource code, dataset, and scripts for reproducing the results in this study.

\section{Simulation studies} \label{SM:sim}

To examine the properties of the proposed model, we fitted three models to each simulated epidemic: a traditional SIR model, a standard BC model, and the proposed mixture model that combines both. For the standard and mixture BC models, we considered three alarm functions: power, threshold, and Hill‑type.

\subsection{Power function alarm}

\begin{figure}[H] \centering 
    \includegraphics[width=0.925\textwidth]{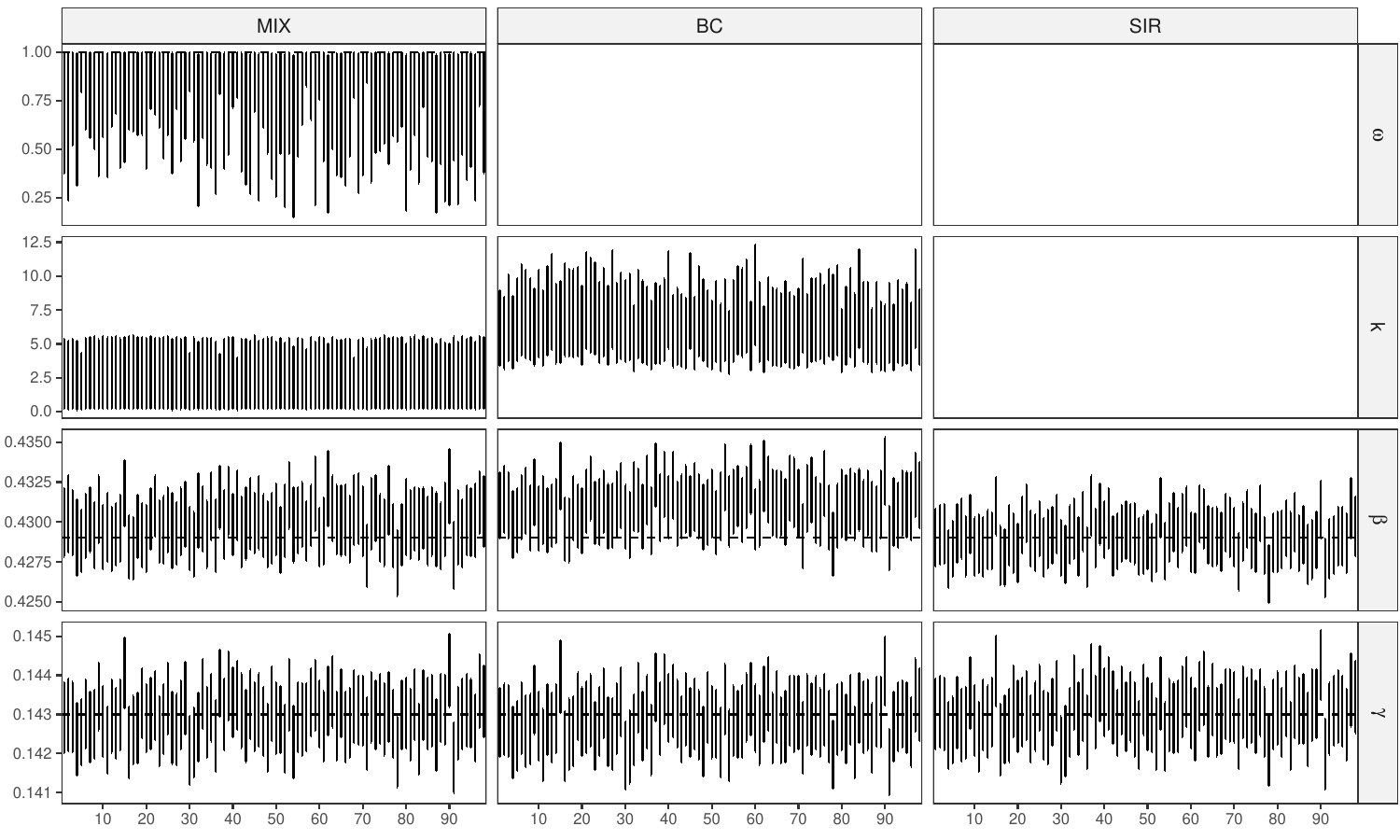}
    \caption{Posterior means and 95\% credible intervals for the parameters in the traditional SIR, BC, and mixture models fitted to data simulated using the SIR model. The dashed line indicates the true value, where applicable.} \label{SM:fig:CovPowerSIR}
\end{figure}

\begin{figure}[H] \centering 
    \includegraphics[width=0.925\textwidth]{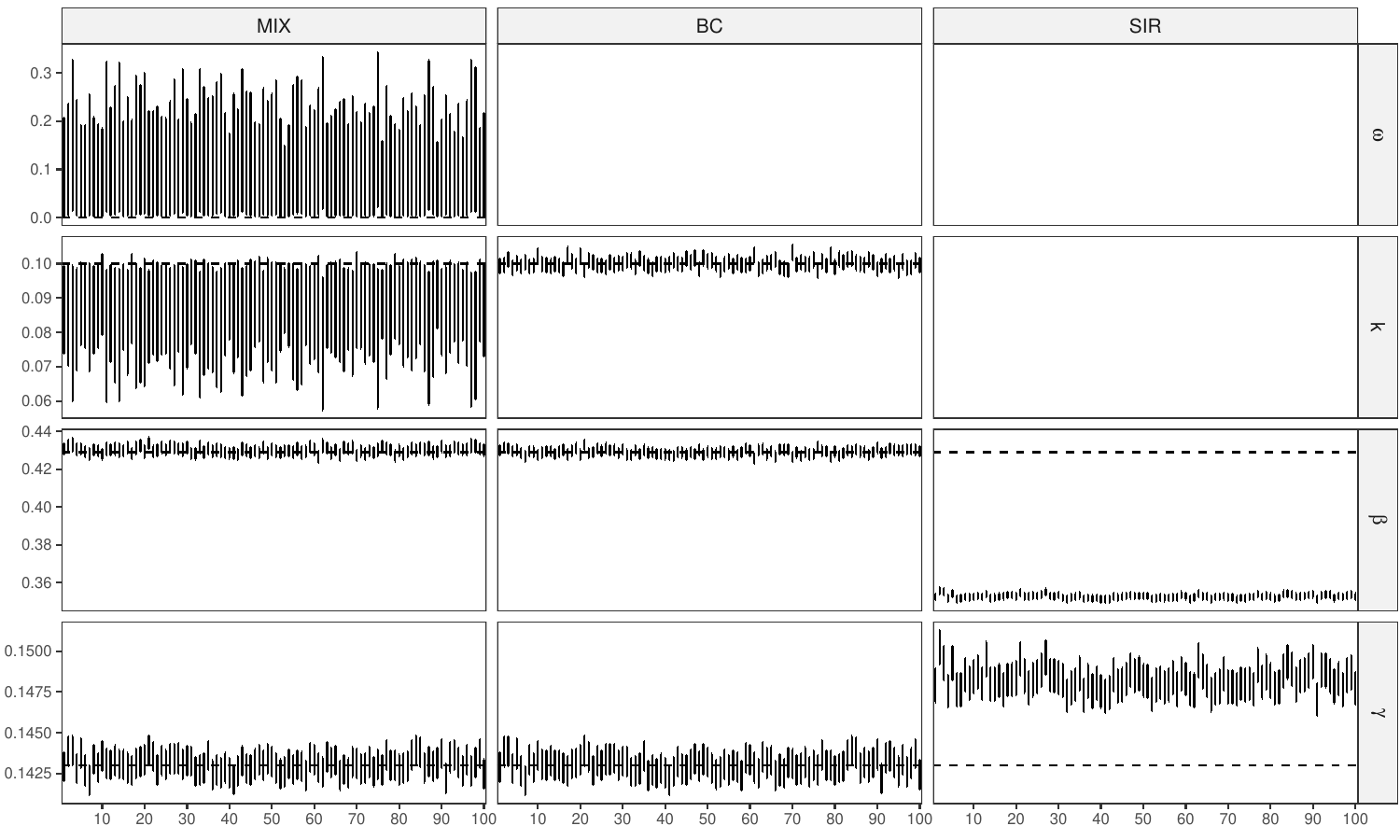}
    \caption{Posterior means and 95\% credible intervals for the parameters in the traditional SIR, BC, and mixture models fitted to data simulated using the BC model. The dashed line indicates the true value, where applicable.} \label{SM:fig:CovPowerBC}
\end{figure}

\begin{figure}[H] \centering 
    \includegraphics[width=0.925\textwidth]{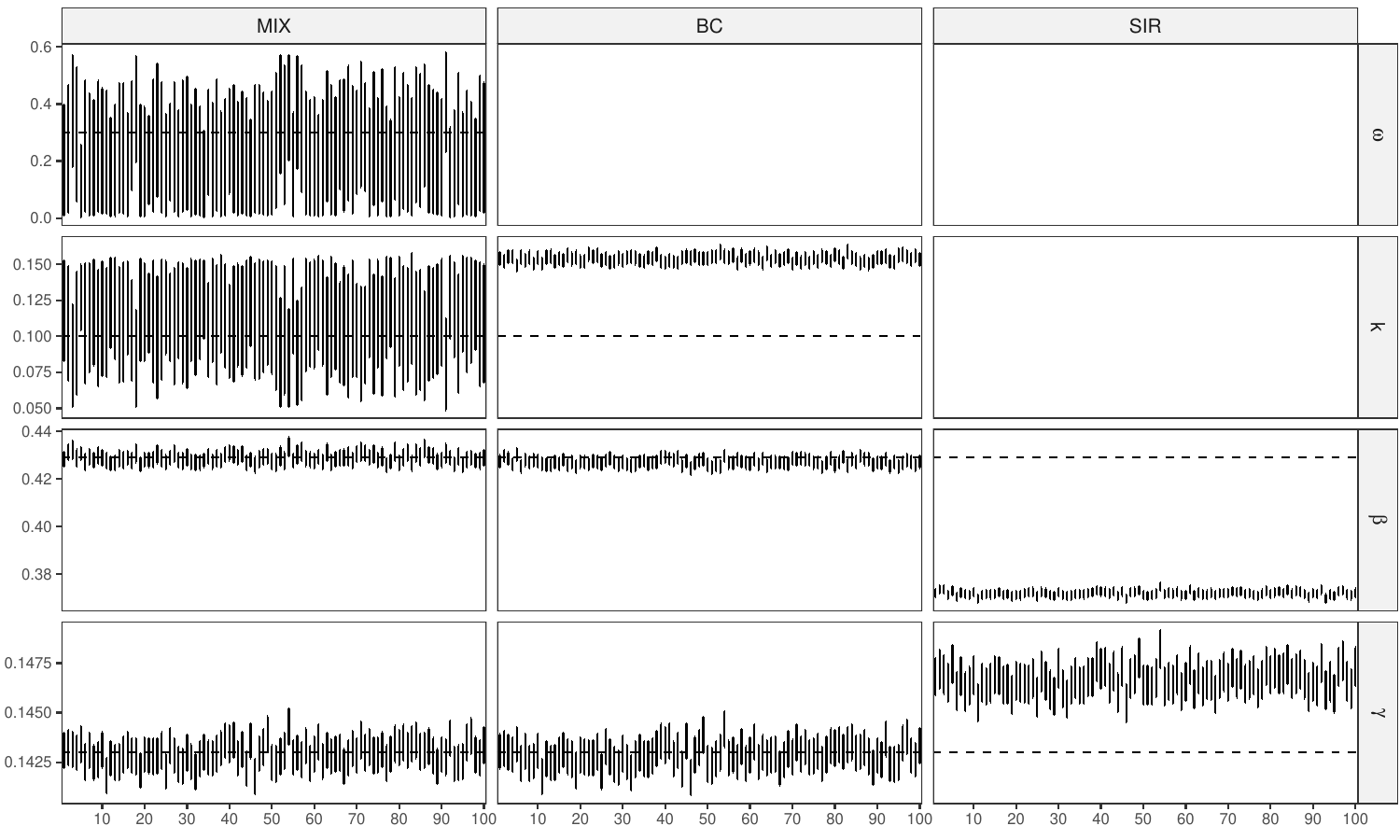}
    \caption{Posterior means and 95\% credible intervals for the parameters in the traditional SIR, BC, and mixture models fitted to data simulated using the mixture model with 30\% risk-neutral and 70\% risk-averse individual. The dashed line indicates the true value, where applicable.} \label{SM:fig:CovPowerMIX30}
\end{figure}

\begin{figure}[H] \centering 
    \includegraphics[width=0.925\textwidth]{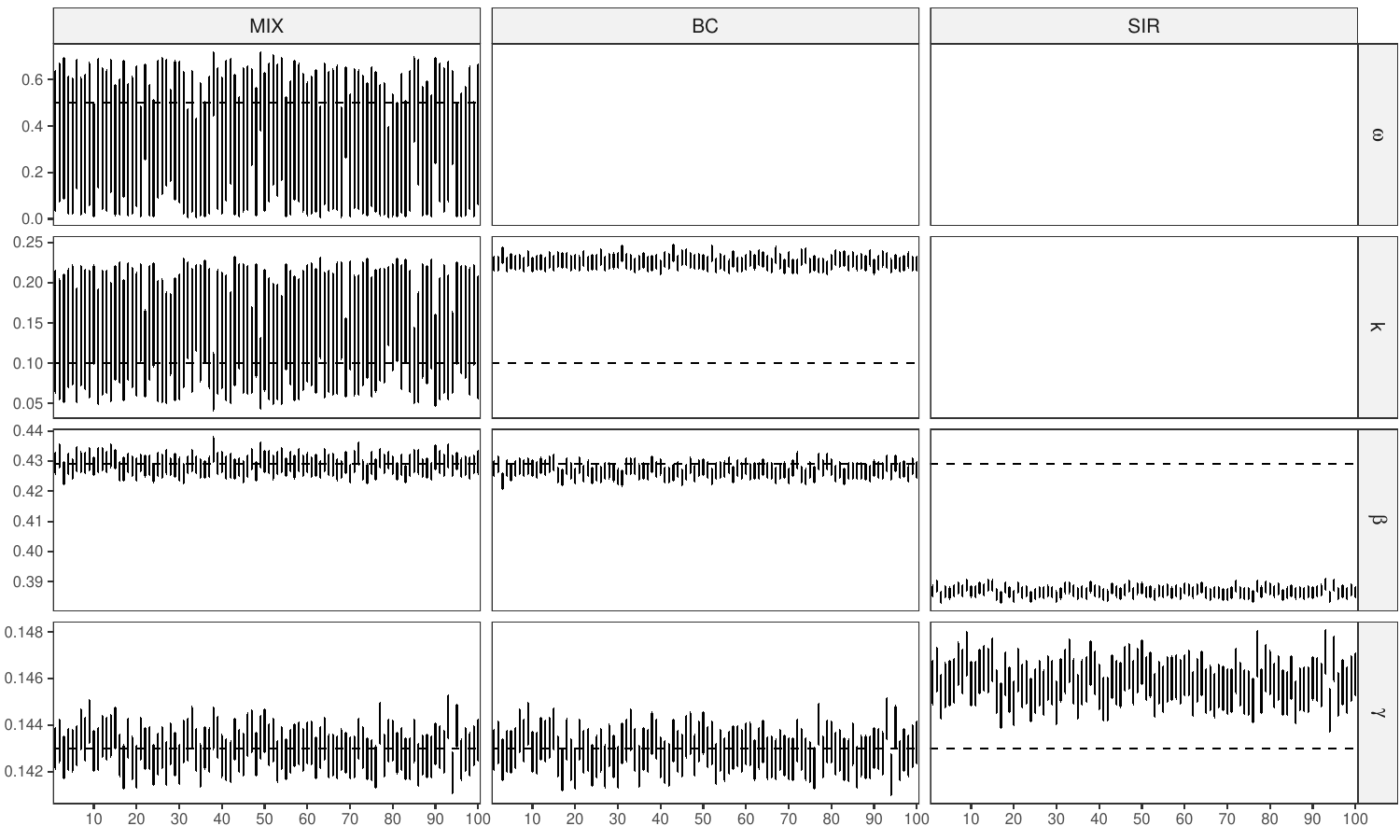}
    \caption{Posterior means and 95\% credible intervals for the parameters in the traditional SIR, BC, and mixture models fitted to data simulated using the mixture model with 50\% risk-neutral and 50\% risk-averse individual. The dashed line indicates the true value, where applicable.} \label{SM:fig:CovPowerMIX50}
\end{figure}

\begin{figure}[H] \centering 
    \includegraphics[width=0.925\textwidth]{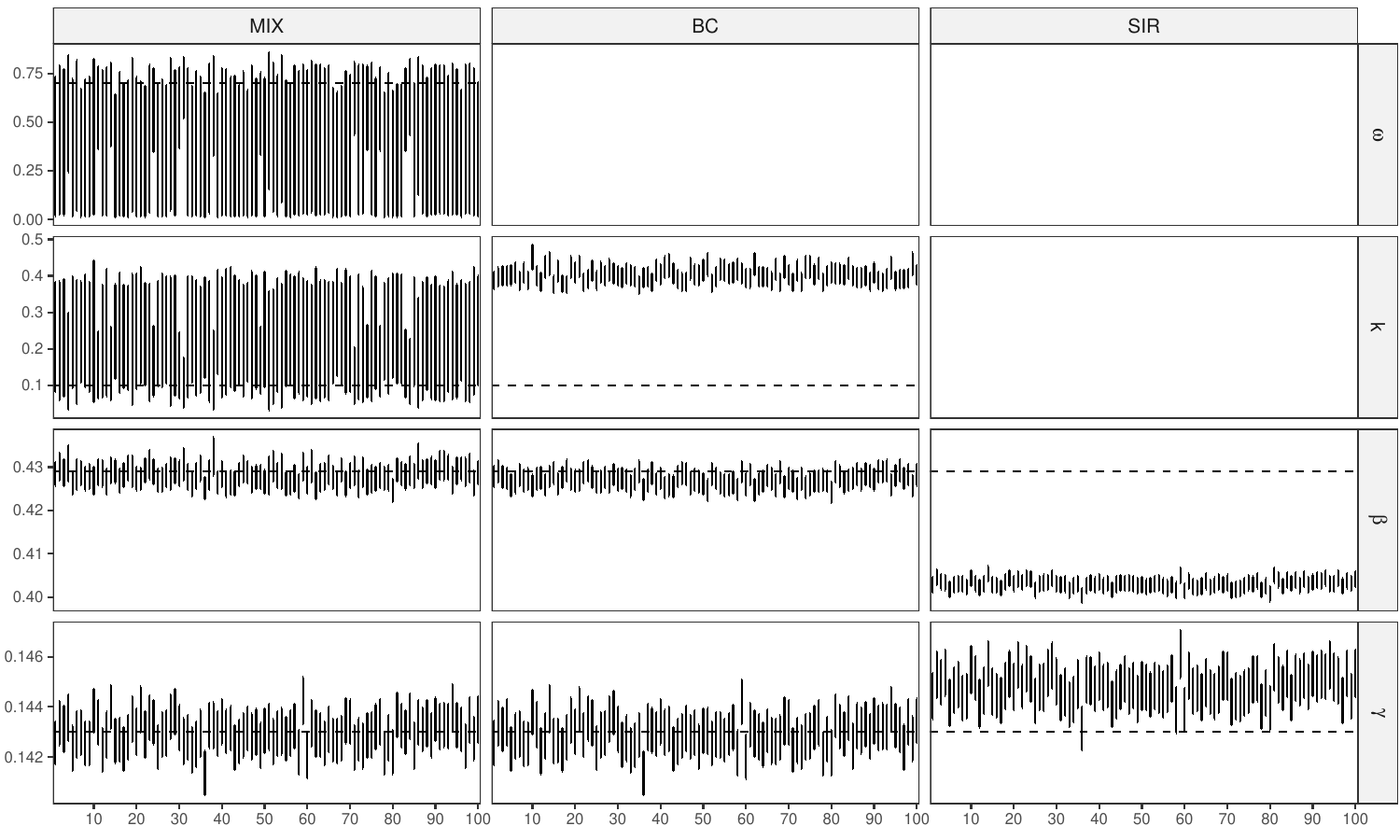}
    \caption{Posterior means and 95\% credible intervals for the parameters in the traditional SIR, BC, and mixture models fitted to data simulated using the mixture model with 70\% risk-neutral and 30\% risk-averse individual. The dashed line indicates the true value, where applicable.} \label{SM:fig:CovPowerMIX70}
\end{figure}

\begin{figure}[H] \centering 
    \includegraphics[width=\textwidth]{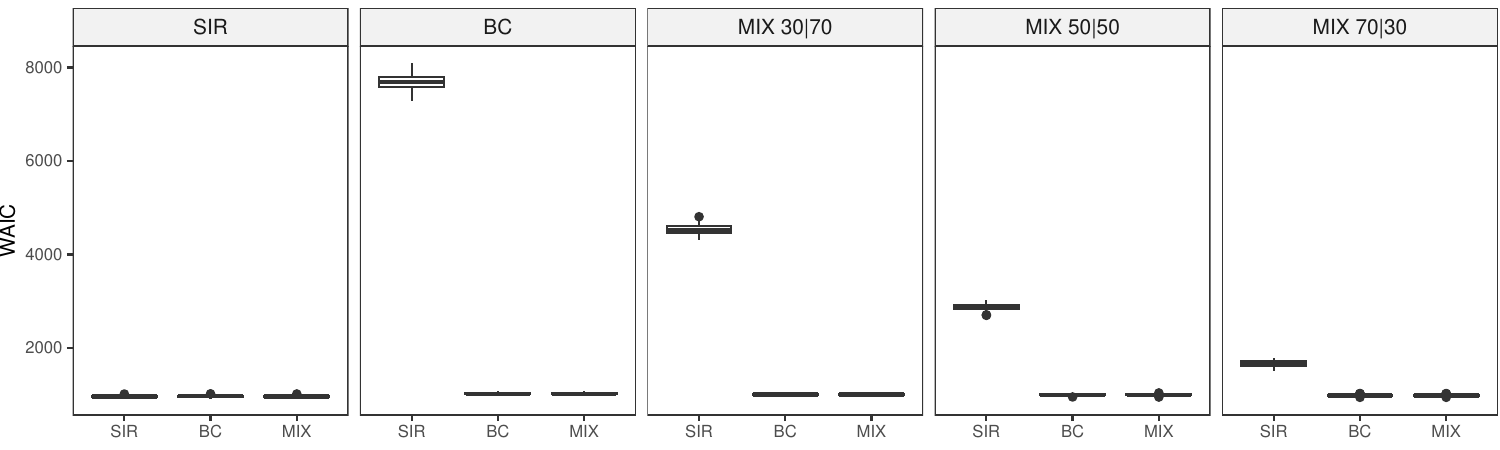}
    \caption{WAIC values obtained by fitting the traditional SIR, BC, and mixture models to data simulated from different mechanisms (SIR, BC, MIX 30$|$70, MIX 50$|$50, and MIX 70$|$30). Lower WAIC values indicate better fit. } \label{SM:fig:WAICPower}
\end{figure}

The complete discussion of models using the power alarm function is provided in Section 4 of the main manuscript. This section presents only complementary results that support that discussion. Figures~\ref{SM:fig:CovPowerSIR}–\ref{SM:fig:CovPowerMIX70} display posterior means and 95\% credible intervals for the parameters of the traditional SIR, BC, and mixture models fitted to data simulated under the five scenarios. Table~\ref{SM:tab:PowerSummary} reports posterior summaries and coverage rates for the parameters estimated by fitting the mixture models to all simulated datasets. Finally, Figure~\ref{SM:fig:WAICPower} shows the widely applicable information criterion (WAIC) values used to compare the performance of the fitted models.

\begin{table}[H]
\caption{Parameters estimated by the mixture models for data simulated from different mechanisms (SIR, BC, MIX 30$|$70, MIX 50$|$50, and MIX 70$|$30). Results are presented as posterior mean (standard deviation) $|$ Coverage rate. Coverage rate was calculated based on 95\% credible intervals. Values are averaged over the 100 generated datasets. \label{SM:tab:PowerSummary} }%
\begin{tabular*}{\columnwidth}{@{\extracolsep\fill}lcccccc@{\extracolsep\fill}}
\toprule
 & $\omega$ & $k$ & $\gamma$ & $\beta$ \\ 
\midrule
SIR         & 0.862 (0.140) $|$ 0.74 & 1.911 (1.360) $|$ ~~~-~~ & 0.143 (0.000) $|$ 0.95 & 0.430 (0.001) $|$ 0.90 \\
BC          & 0.096 (0.065) $|$ 0.73 & 0.088 (0.008) $|$ 0.43 & 0.143 (0.000) $|$ 0.98 & 0.430 (0.002) $|$ 0.94 \\
MIX $30|70$ & 0.241 (0.113) $|$ 0.99 & 0.111 (0.020) $|$ 0.99 & 0.143 (0.000) $|$ 0.93 & 0.429 (0.002) $|$ 0.96 \\
MIX $50|50$ & 0.373 (0.152) $|$ 0.94 & 0.132 (0.039) $|$ 0.92 & 0.143 (0.000) $|$ 0.94 & 0.429 (0.002) $|$ 0.95 \\
MIX $70|30$ & 0.423 (0.201) $|$ 0.84 & 0.220 (0.087) $|$ 0.81 & 0.143 (0.000) $|$ 0.96 & 0.428 (0.002) $|$ 0.89 \\
\bottomrule
\end{tabular*}
\begin{tablenotes}%
\item Mean and SD are averaged over all the generated datasets.
\end{tablenotes}
\end{table}

\subsection{Threshold function alarm}

\begin{figure}[H] \centering 
    \includegraphics[width=\textwidth]{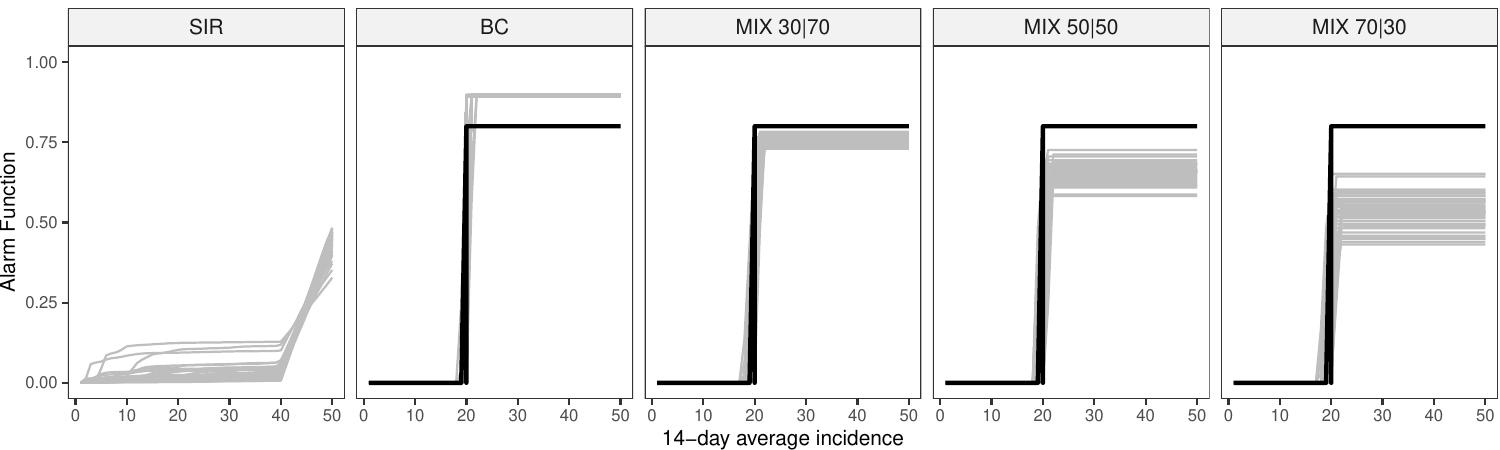}
    \caption{True alarm function (black lines) and posterior mean estimates of alarm functions (gray lines) estimated by the mixture model. Results for 50 simulated epidemics using the traditional SIR model (SIR), behavioral change model (BC), and mixture models with different proportions of risk-neutral populations: 30\%, 50\%, and 70\% (MIX 30$|$70, MIX 50$|$50, and MIX 70$|$30, respectively).}
    \label{fig:ThreshAlarm}
\end{figure}

Both the standard SIR and BC models recover their parameters when the data are simulated directly from them. Otherwise, they fail to estimate the parameters correctly. The mixture model recovers parameters when the data are simulated from itself under any setting, but it has some difficulty estimating $\omega$ when the data come from other models. Unlike the results obtained with the power alarm function, when using the threshold alarm function, the mixture model does not appear to learn about $\omega$, yielding wide 95\% credible intervals. Indeed, the results suggest a potential nonidentifiability issue between $\omega$ and $\delta$. When examining the estimated alarm function (Figure~\ref{fig:ThreshAlarm}), it is evident that the mixture model can identify the change point but struggles to estimate the corresponding level of alarm in the population.

\begin{figure}[H] \centering 
    \includegraphics[width=0.925\textwidth]{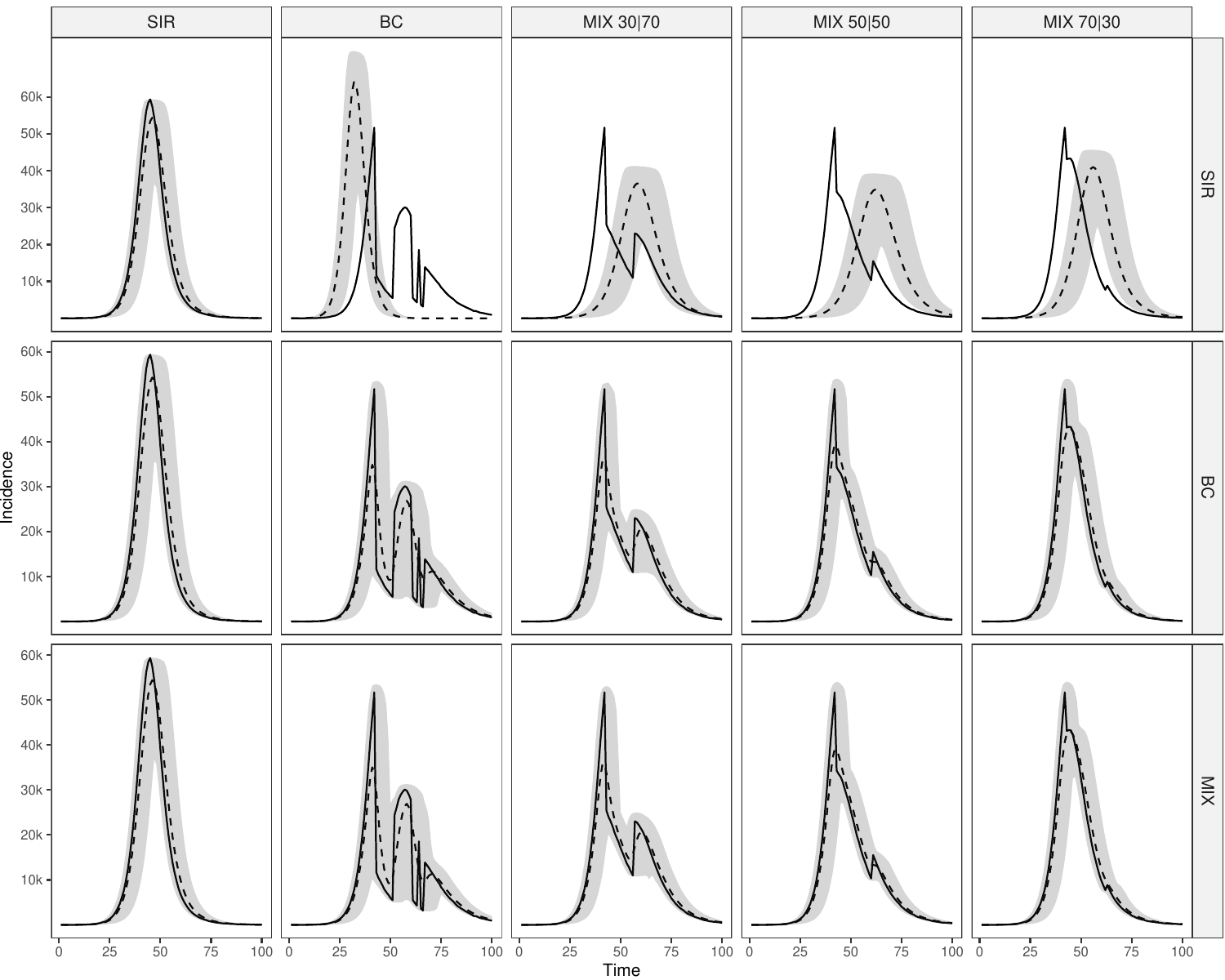}
    \caption{True incidence curve (solid line), posterior predictive mean (dashed, dotted, long-dashed lines for SIR, BC, and mixture models, respectively), and 95\% credible intervals for a randomly selected simulation from different mechanisms (SIR, BC, MIX 30$|$70, MIX 50$|$50, and MIX 70$|$30). BC and MIX curves are overlapping.}
    \label{fig:ThreshEst}
\end{figure}

\begin{figure}[H] \centering 
    \includegraphics[width=0.925\textwidth]{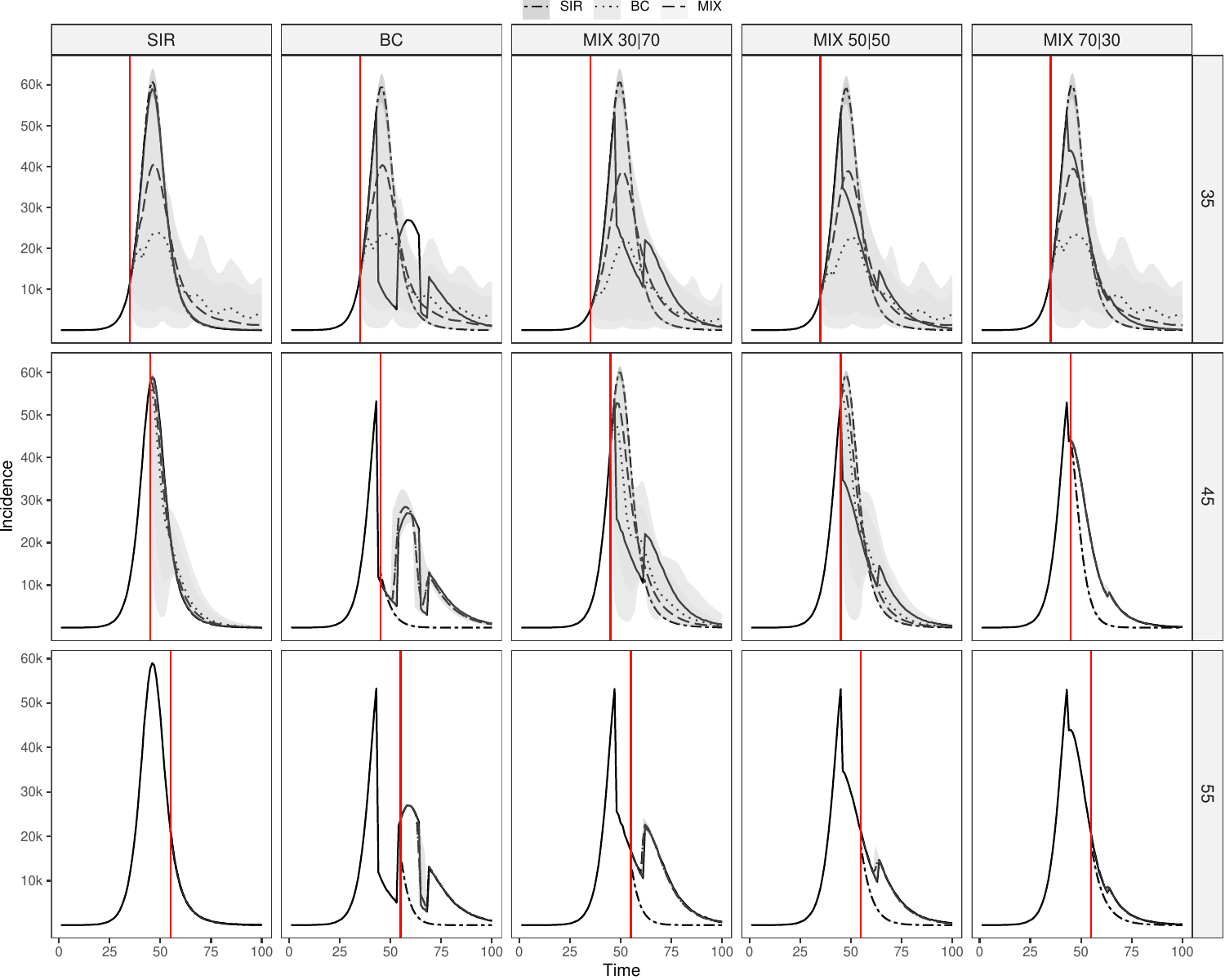}
    \caption{Posterior predictive mean (dashed, dotted, and long-dashed lines for SIR, BC, and mixture models, respectively) and 95\% credible intervals for forecasts of future incidence compared to the true values (solid line) for a randomly selected simulation. Results were generated considering different cutoffs at days 35, 45, and 55. BC and MIX curves are overlapping.}
    \label{fig:ThreshFor}
\end{figure}

Despite the difficulty in estimating parameters in scenarios where the data were not simulated from the mixture model itself, the mixture model with a threshold alarm function still provides a satisfactory reconstruction of the epidemic incidence curve. As shown in Figure~\ref{fig:ThreshEst}, both the standard and mixture BC models adapt effectively to all incidence curve shapes. In terms of forecasting (Figure~\ref{fig:ThreshFor}), the SIR model overestimates the epidemic peak, leading to a faster decline in the predicted epidemic curve compared to the true trajectory, which makes it unable to capture the non‑constant post‑peak behavior. Both the standard and mixture BC models fail to provide accurate predictions when the cutoff point is very early (day 35). However, their performance improves when the cutoff occurs later (day 45 or 55). When comparing the two, the mixture model is slightly more precise, yielding narrower predictive intervals.

\begin{figure}[H] \centering 
    \includegraphics[width=0.925\textwidth]{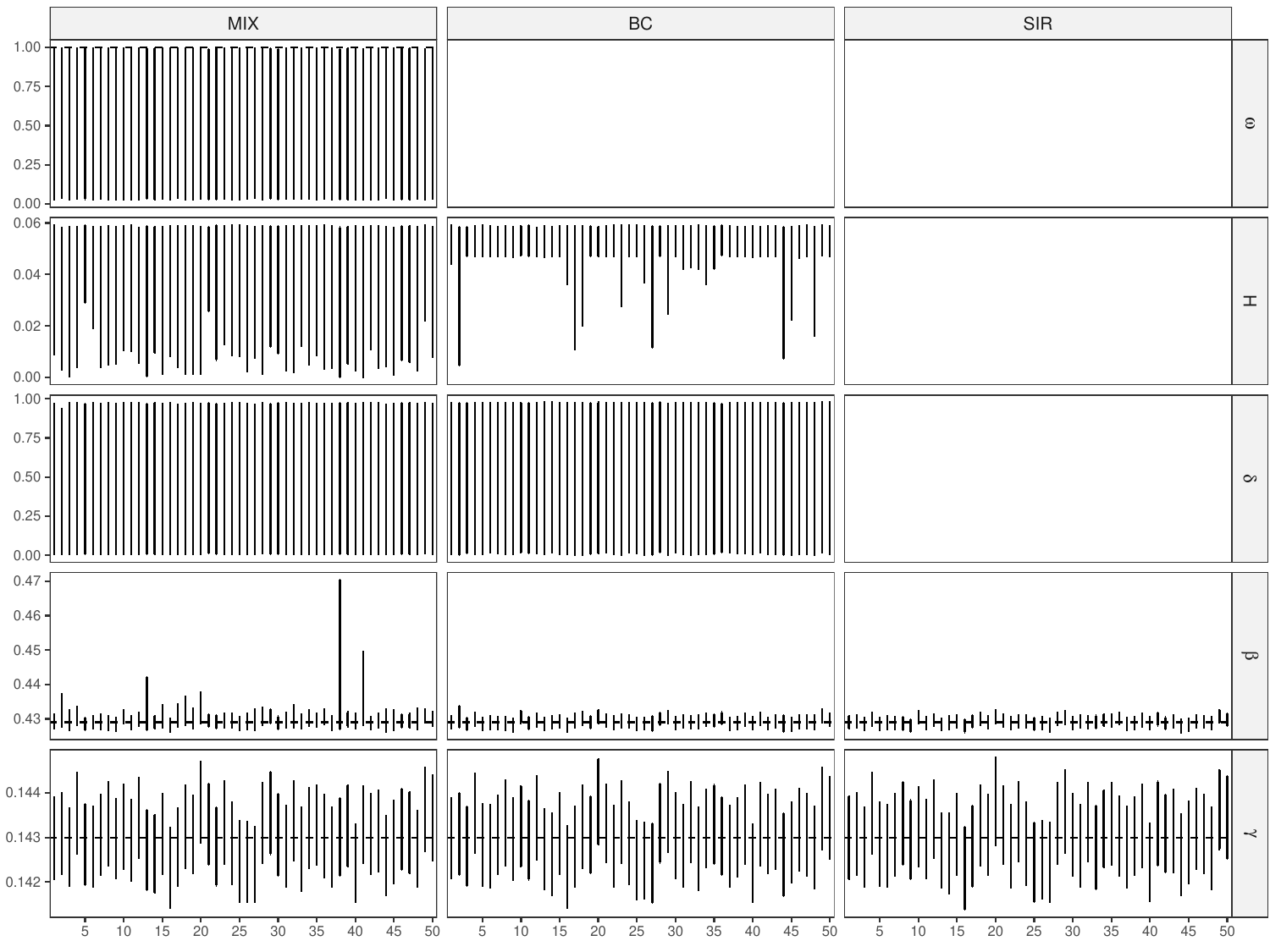}
    \caption{Posterior means and 95\% credible intervals for the parameters in the traditional SIR, BC, and mixture models fitted to data simulated using the SIR model. The dashed line indicates the true value, where applicable.} \label{SM:fig:CovThreshSIR}
\end{figure}

\begin{figure}[H] \centering 
    \includegraphics[width=0.925\textwidth]{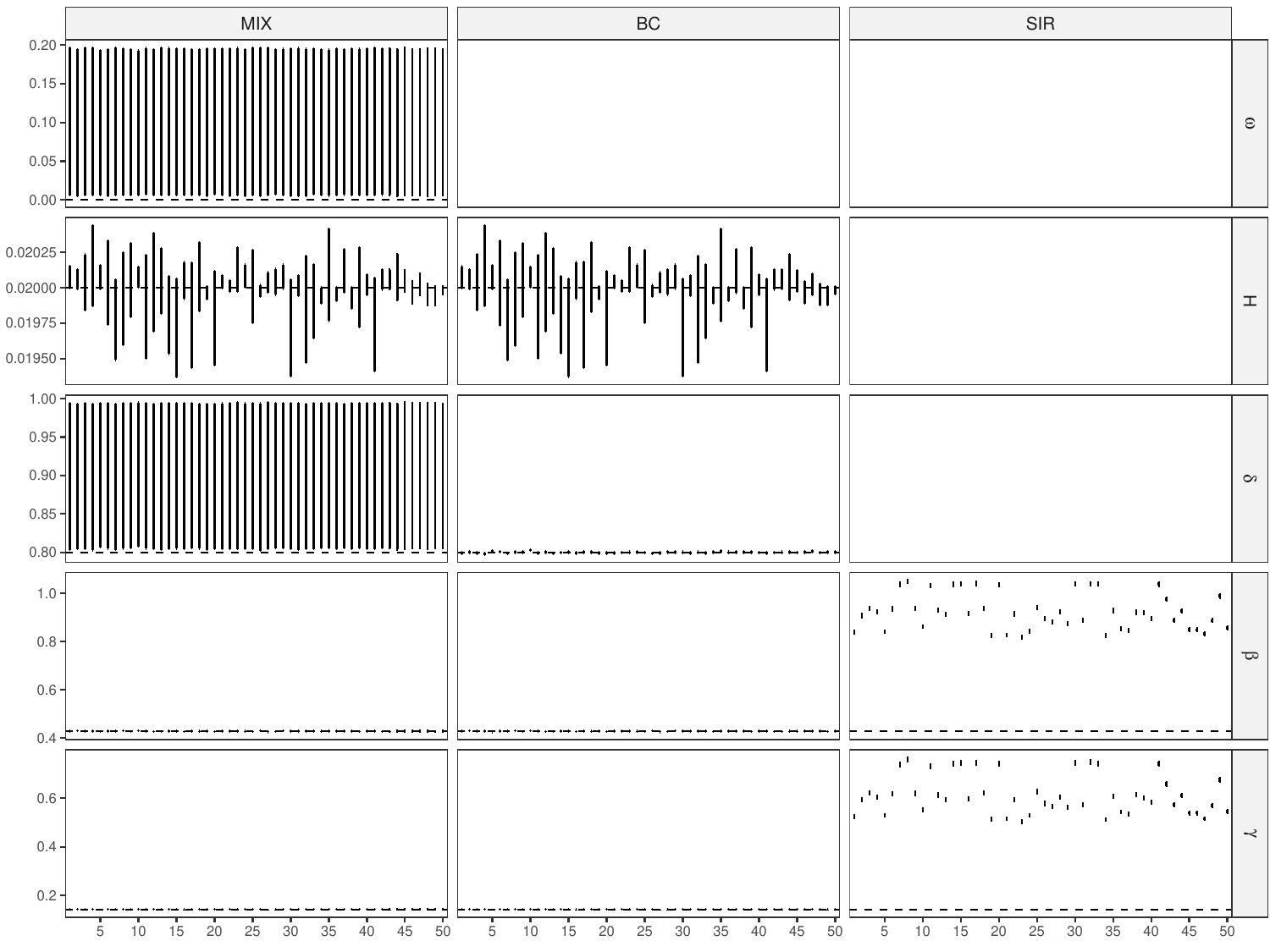}
    \caption{Posterior means and 95\% credible intervals for the parameters in the traditional SIR, BC, and mixture models fitted to data simulated using the BC model. The dashed line indicates the true value, where applicable.} \label{SM:fig:CovThreshBC}
\end{figure}

\begin{figure}[H] \centering 
    \includegraphics[width=0.925\textwidth]{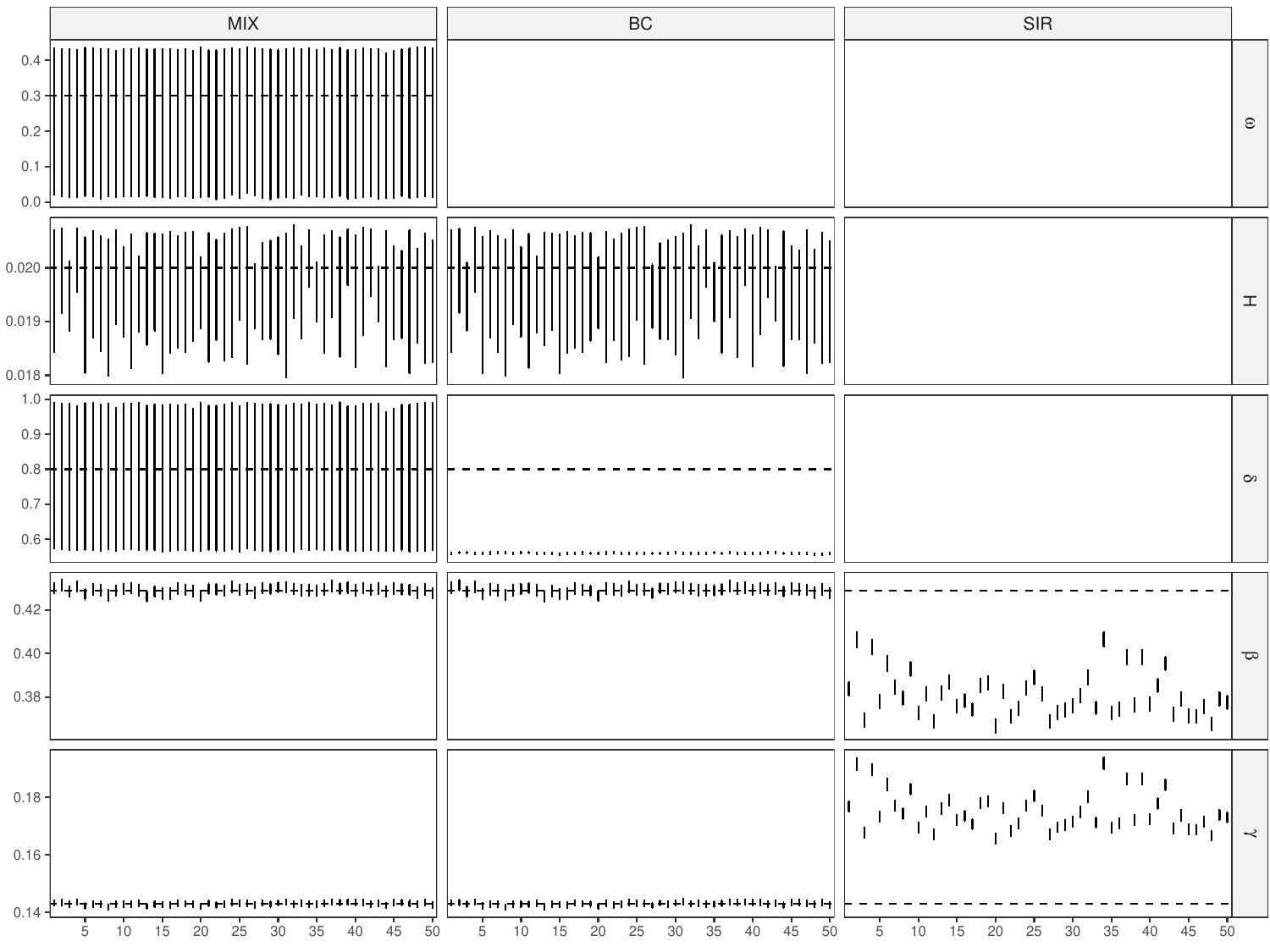}
    \caption{Posterior means and 95\% credible intervals for the parameters in the traditional SIR, BC, and mixture models fitted to data simulated using the mixture model with 30\% risk-neutral and 70\% risk-averse individual. The dashed line indicates the true value, where applicable.}  \label{SM:fig:CovThreshMIX30}
\end{figure}

\begin{figure}[H] \centering 
    \includegraphics[width=0.925\textwidth]{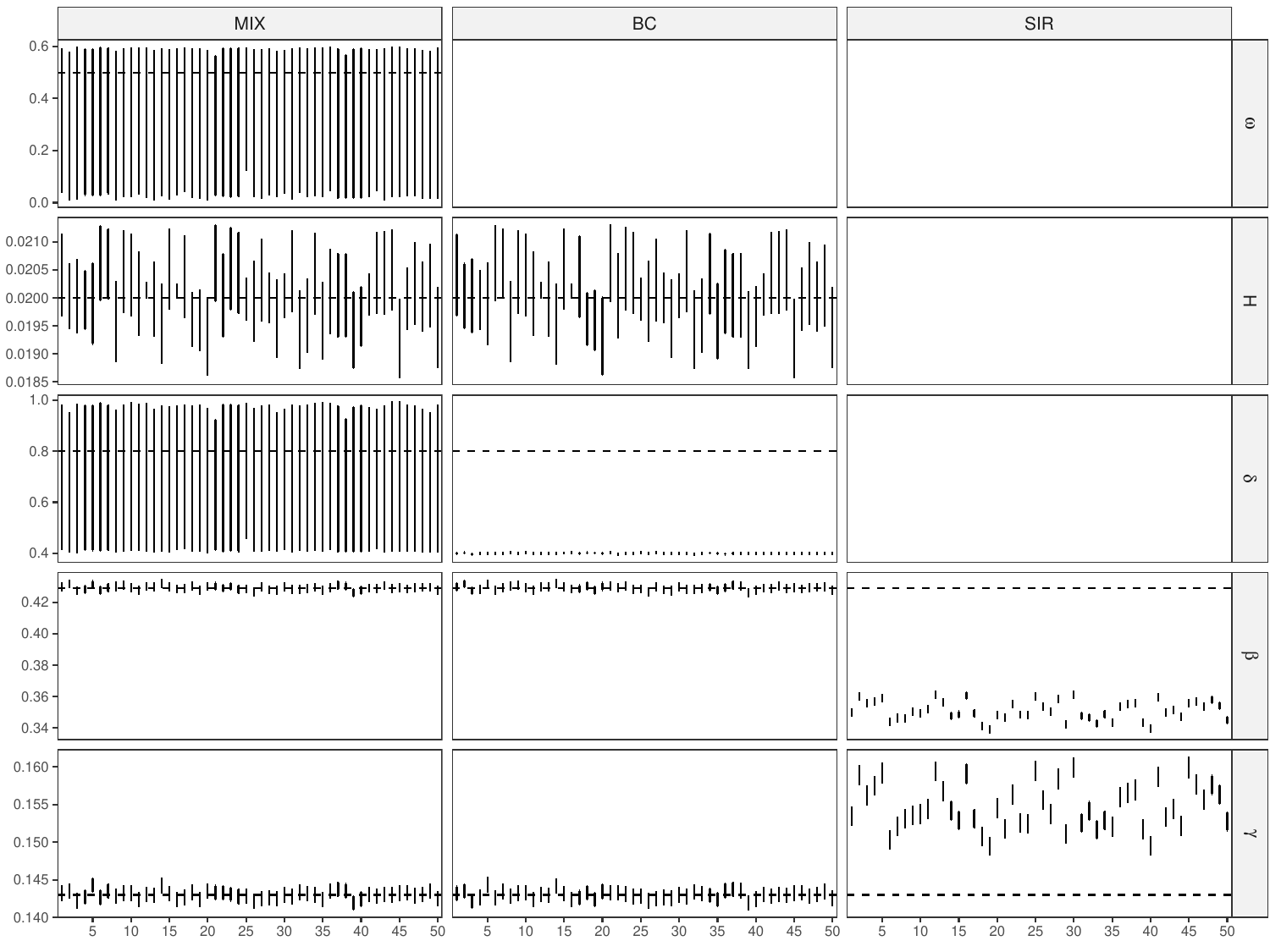}
    \caption{Posterior means and 95\% credible intervals for the parameters in the traditional SIR, BC, and mixture models fitted to data simulated using the mixture model with 50\% risk-neutral and 50\% risk-averse individual. The dashed line indicates the true value, where applicable.} \label{SM:fig:CovThreshMIX50}
\end{figure}

\begin{figure}[H] \centering 
    \includegraphics[width=0.925\textwidth]{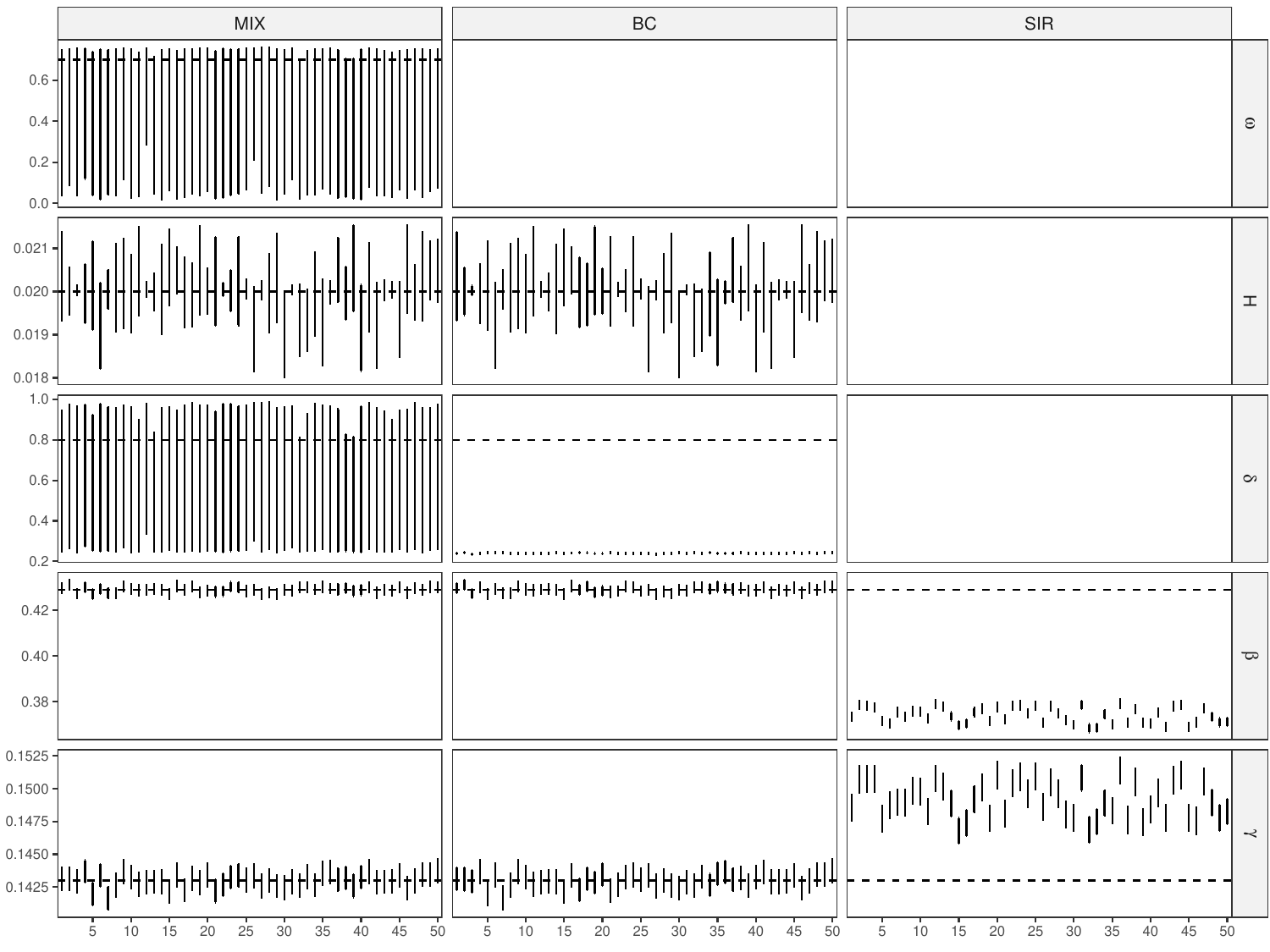}
    \caption{Posterior means and 95\% credible intervals for the parameters in the traditional SIR, BC, and mixture models fitted to data simulated using the mixture model with 70\% risk-neutral and 30\% risk-averse individual. The dashed line indicates the true value, where applicable.}   \label{SM:fig:CovThreshMIX70}
\end{figure}

\begin{figure}[H] \centering 
    \includegraphics[width=\textwidth]{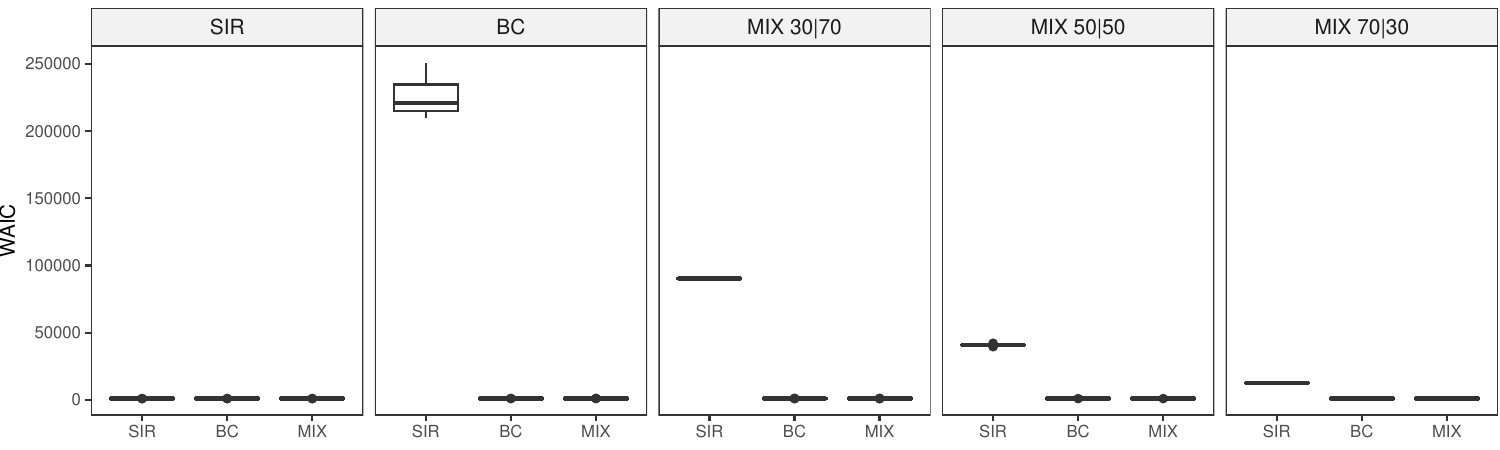}
    \caption{WAIC values obtained by fitting the traditional SIR, BC, and mixture models to data simulated from different mechanisms (SIR, BC, MIX 30$|$70, MIX 50$|$50, and MIX 70$|$30). Lower WAIC values indicate better fit. } \label{SM:fig:WAICThresh}
\end{figure}

\subsection{Hill-type function alarm}

\begin{figure}[H] \centering 
    \includegraphics[width=\textwidth]{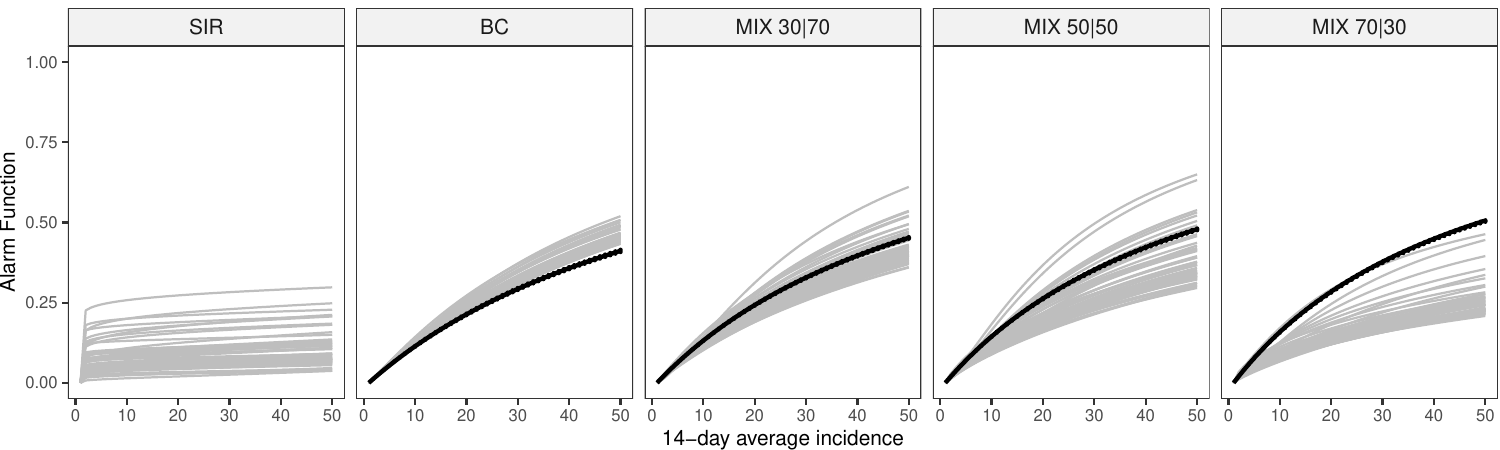}
    \caption{True alarm function (black lines) and posterior mean estimates of alarm functions (gray lines) estimated by the mixture model. Results for 89 simulated epidemics using the traditional SIR model (SIR), behavioral change model (BC), and mixture models with different proportions of risk-neutral populations: 30\%, 50\%, and 70\% (MIX 30$|$70, MIX 50$|$50, and MIX 70$|$30, respectively).}
    \label{fig:Hill2pAlarm}
\end{figure}

When the data are generated from their own structure, both the SIR and BC models successfully recover their parameters. However, they perform poorly when the data originate from a different model. The mixture model, in contrast, retrieves its parameters reliably across all scenarios in which the data are simulated from itself, though it still struggles to estimate $\omega$ when the data come from alternative models. Regarding the alarm function (Figure~\ref{fig:Hill2pAlarm}), the mixture model recovers it satisfactorily, with precision improving as the proportion of risk‑averse individuals increases. This is the same pattern observed when using the power alarm function. 

Even with these parameter‑estimation challenges in settings where the data do not arise from the mixture model, the mixture model with a Hill‑type alarm function still reconstructs the epidemic incidence curve reasonably well. As illustrated in Figure~\ref{fig:HillEst}, both the standard BC model and its mixture counterpart adapt to the incidence curve. Forecasting results (Figure~\ref{fig:HillFor}) reveal that the SIR model tends to overestimate the epidemic peak, which leads to a decline that is too rapid and prevents it from capturing the non‑constant post‑peak behavior. The BC models, standard and mixture, perform poorly when the cutoff is very early (day 35), but their predictive accuracy improves when the cutoff occurs later (day 45 or 55). Between the two, the mixture model generally yields slightly narrower predictive intervals, indicating a modest gain in precision.

\begin{figure}[H] \centering 
    \includegraphics[width=\textwidth]{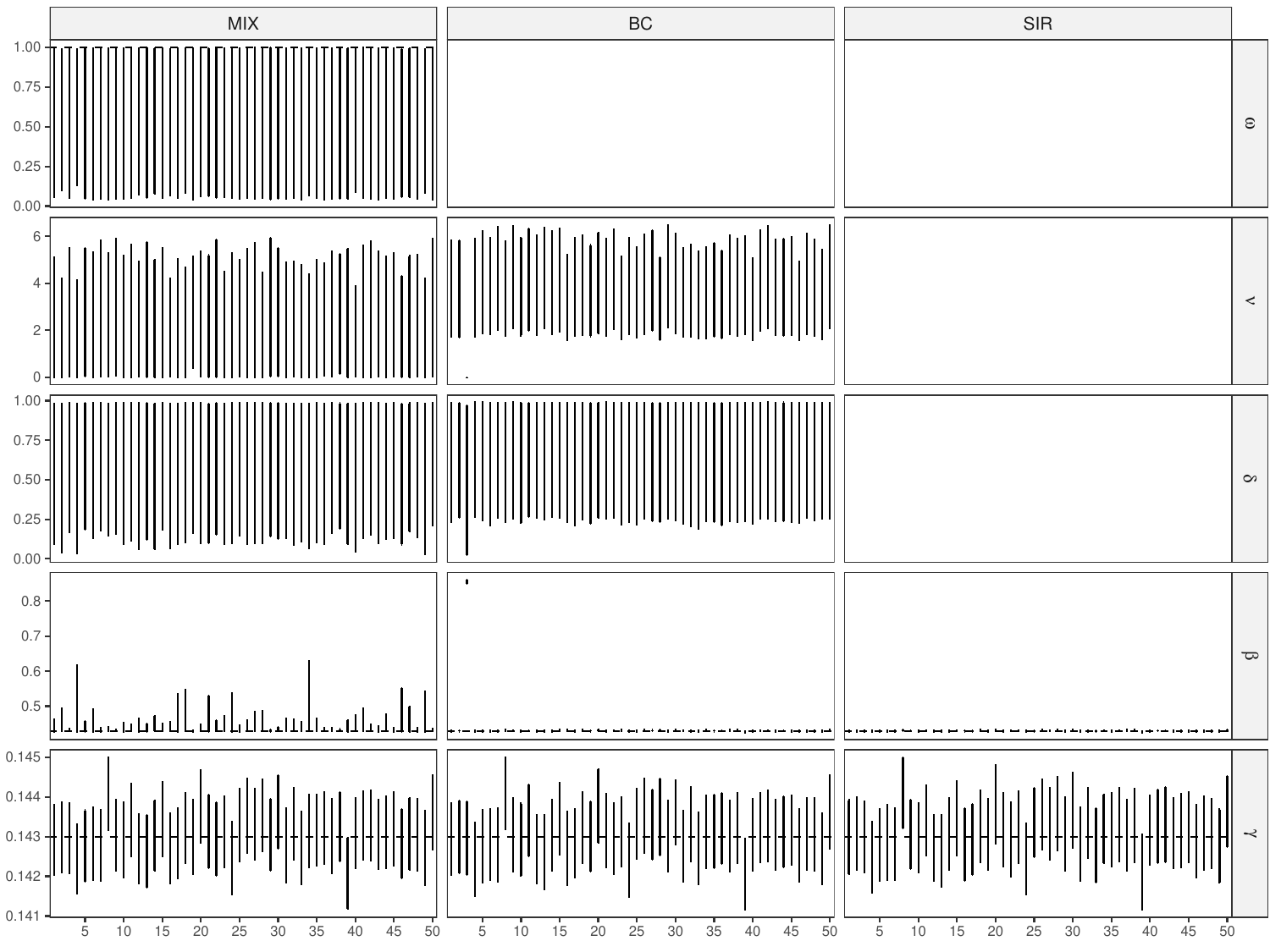}
    \caption{Posterior means and 95\% credible intervals for the parameters in the traditional SIR, BC, and mixture models fitted to data simulated using the SIR model. The dashed line indicates the true value, where applicable.} \label{SM:fig:CovHill2pSIR}
\end{figure}

\begin{figure}[H] \centering 
    \includegraphics[width=\textwidth]{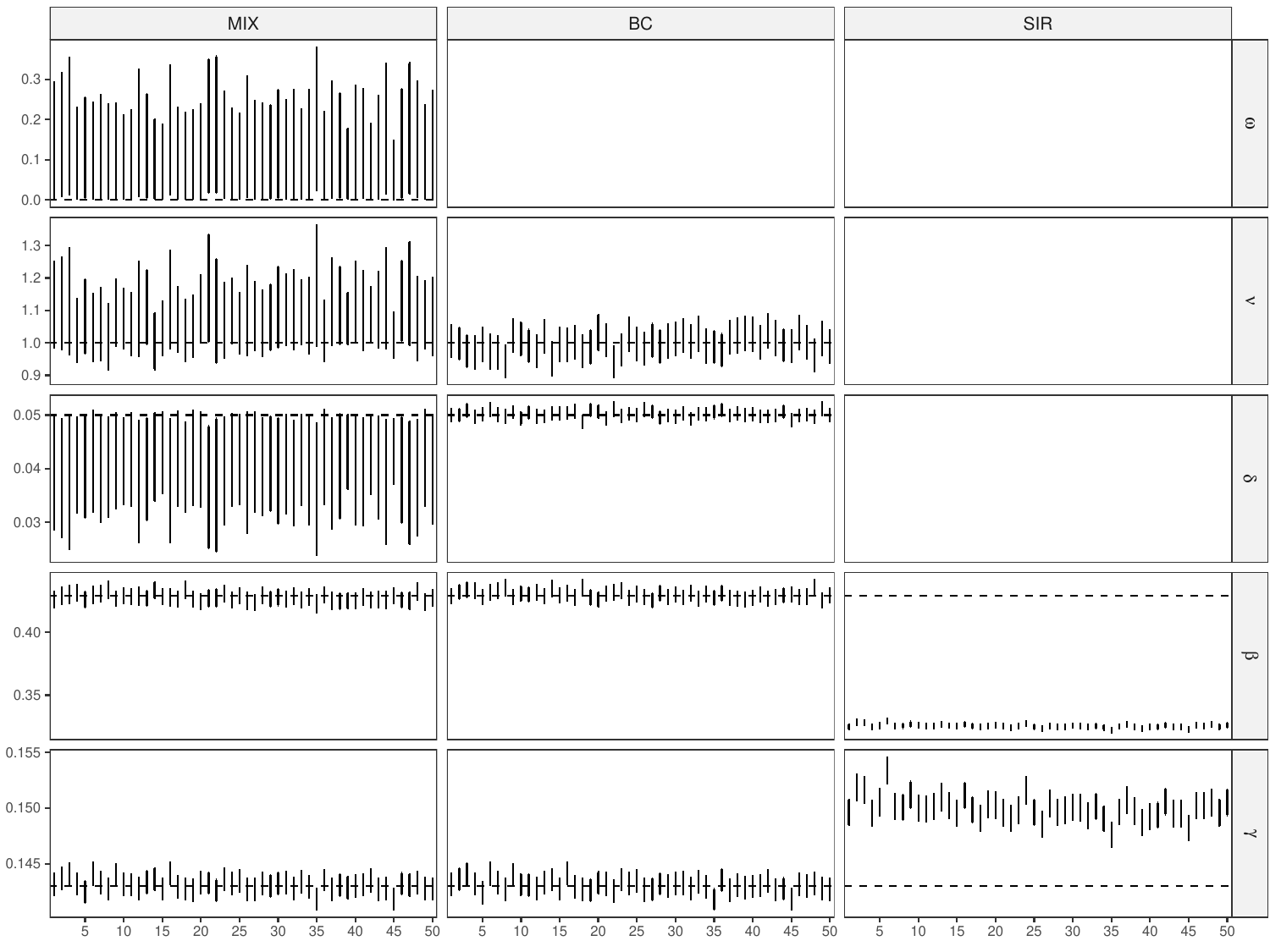}
    \caption{Posterior means and 95\% credible intervals for the parameters in the traditional SIR, BC, and mixture models fitted to data simulated using the BC model. The dashed line indicates the true value, where applicable.} \label{SM:fig:CovHill2pBC}
\end{figure}

\begin{figure}[H] \centering 
    \includegraphics[width=\textwidth]{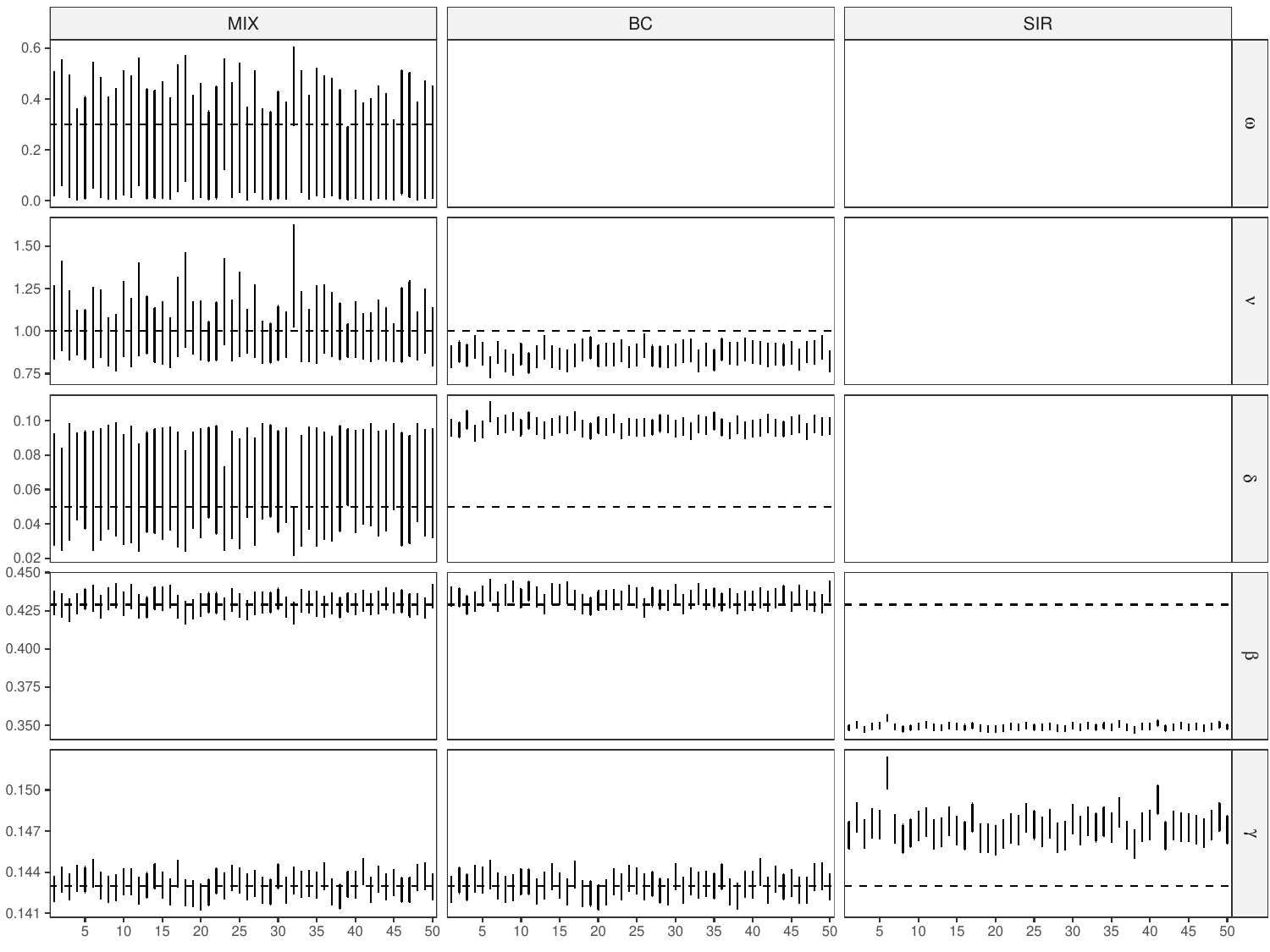}
    \caption{Posterior means and 95\% credible intervals for the parameters in the traditional SIR, BC, and mixture models fitted to data simulated using the mixture model with 30\% risk-neutral and 70\% risk-averse individual. The dashed line indicates the true value, where applicable.}  \label{SM:fig:CovHill2pMIX30}
\end{figure}

\begin{figure}[H] \centering 
    \includegraphics[width=\textwidth]{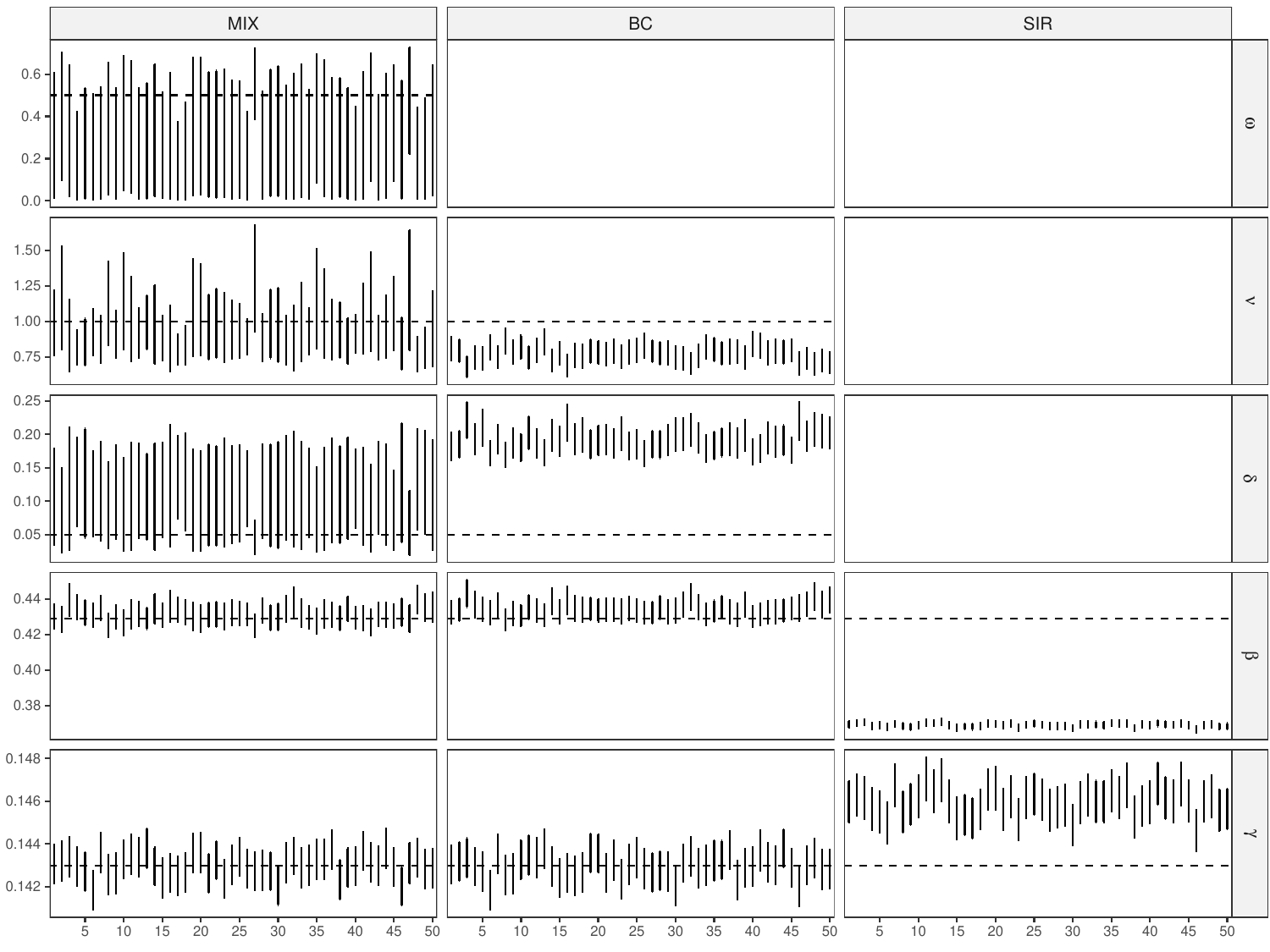}
    \caption{Posterior means and 95\% credible intervals for the parameters in the traditional SIR, BC, and mixture models fitted to data simulated using the mixture model with 50\% risk-neutral and 50\% risk-averse individual. The dashed line indicates the true value, where applicable.} \label{SM:fig:CovHill2pMIX50}
\end{figure}

\begin{figure}[H] \centering 
    \includegraphics[width=\textwidth]{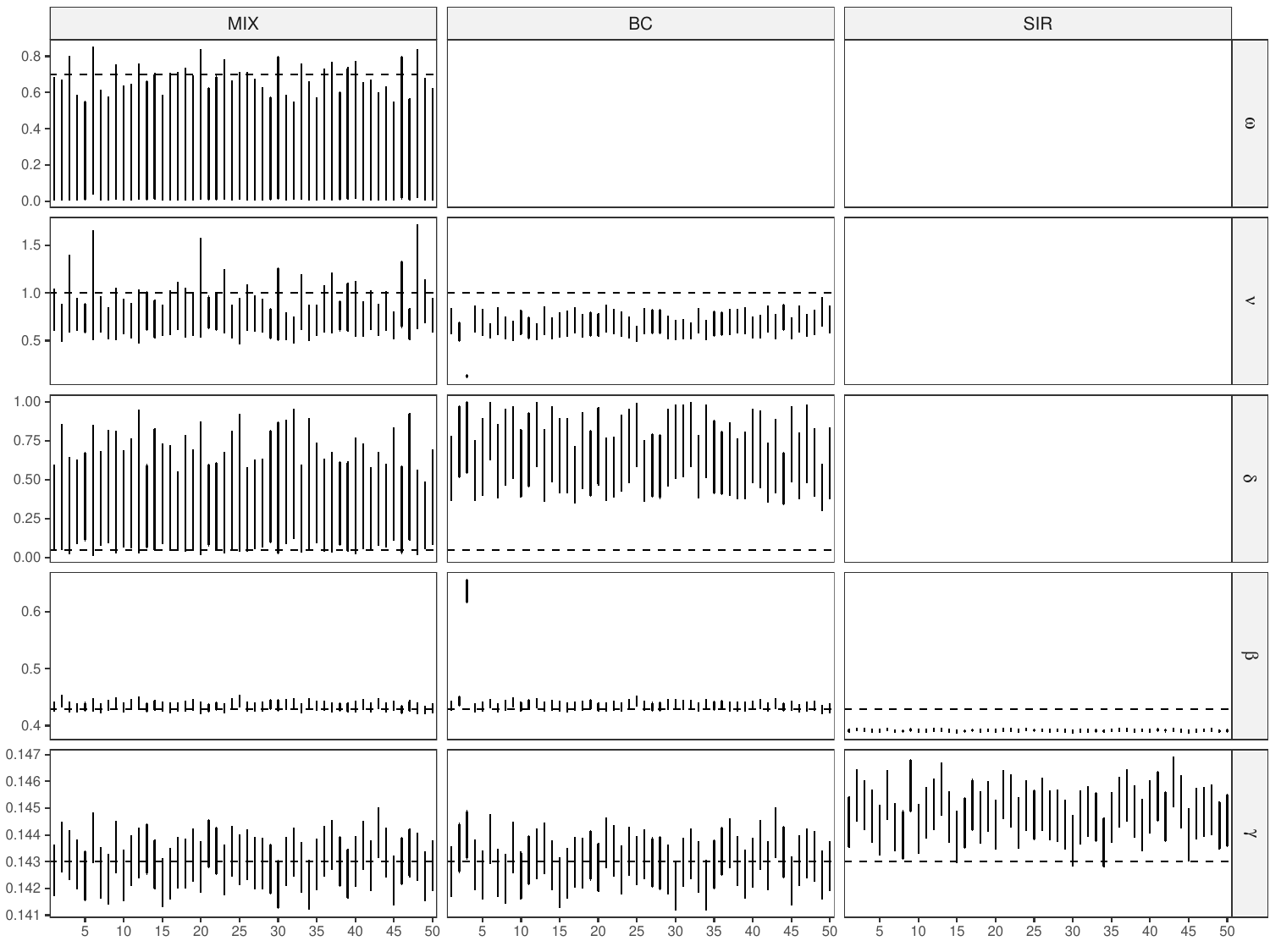}
    \caption{Posterior means and 95\% credible intervals for the parameters in the traditional SIR, BC, and mixture models fitted to data simulated using the mixture model with 70\% risk-neutral and 30\% risk-averse individual. The dashed line indicates the true value, where applicable.}   \label{SM:fig:CovHill2pMIX70}
\end{figure}

\begin{figure}[H] \centering 
    \includegraphics[width=\textwidth]{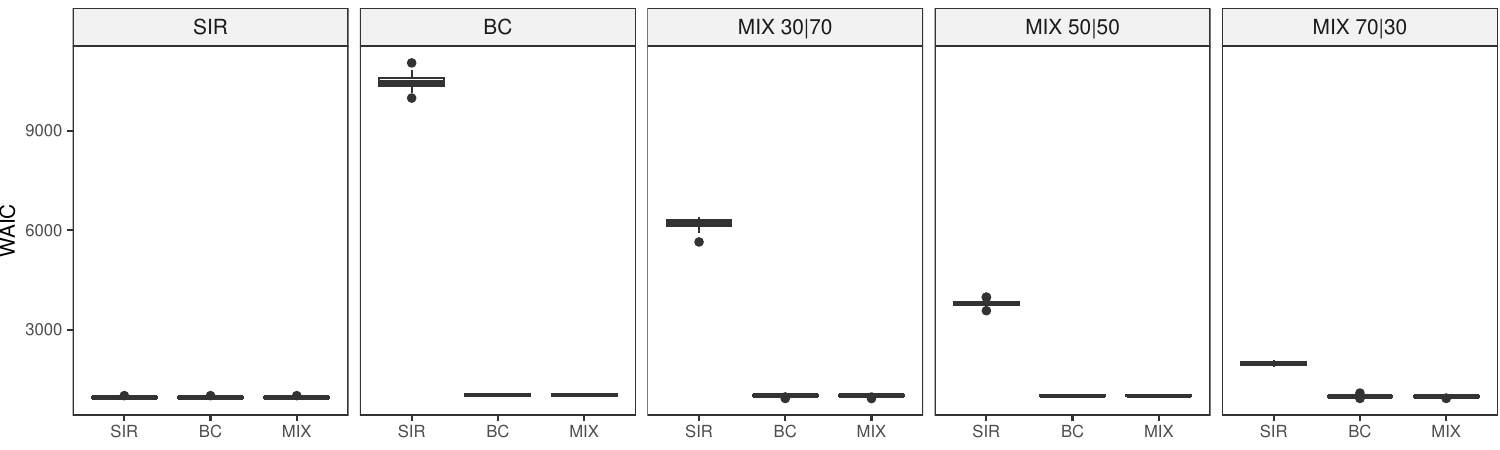}
    \caption{WAIC values obtained by fitting the traditional SIR, BC, and mixture models to data simulated from different mechanisms (SIR, BC, MIX 30$|$70, MIX 50$|$50, and MIX 70$|$30). Lower WAIC values indicate better fit. } \label{SM:fig:WAICHill}
\end{figure}

\begin{figure}[H] \centering 
    \includegraphics[width=\textwidth]{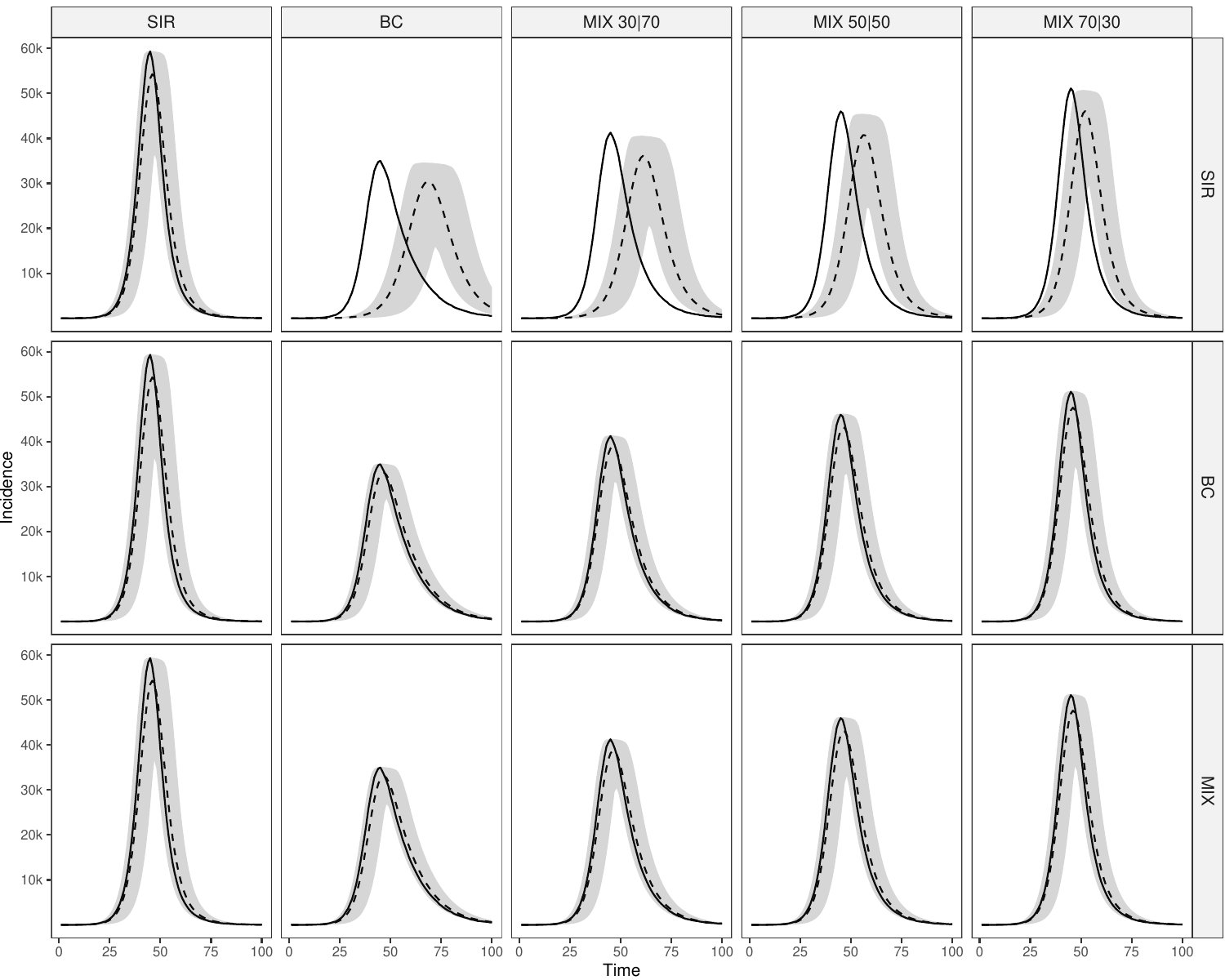}
    \caption{True incidence curve (solid line), posterior predictive mean (dashed, dotted, long-dashed lines for SIR, BC, and mixture models, respectively), and 95\% credible intervals for a randomly selected simulation from different mechanisms (SIR, BC, MIX 30$|$70, MIX 50$|$50, and MIX 70$|$30). BC and MIX curves are overlapping.}
    \label{fig:Hill2pEst}
\end{figure}

\begin{figure}[H] \centering 
    \includegraphics[width=\textwidth]{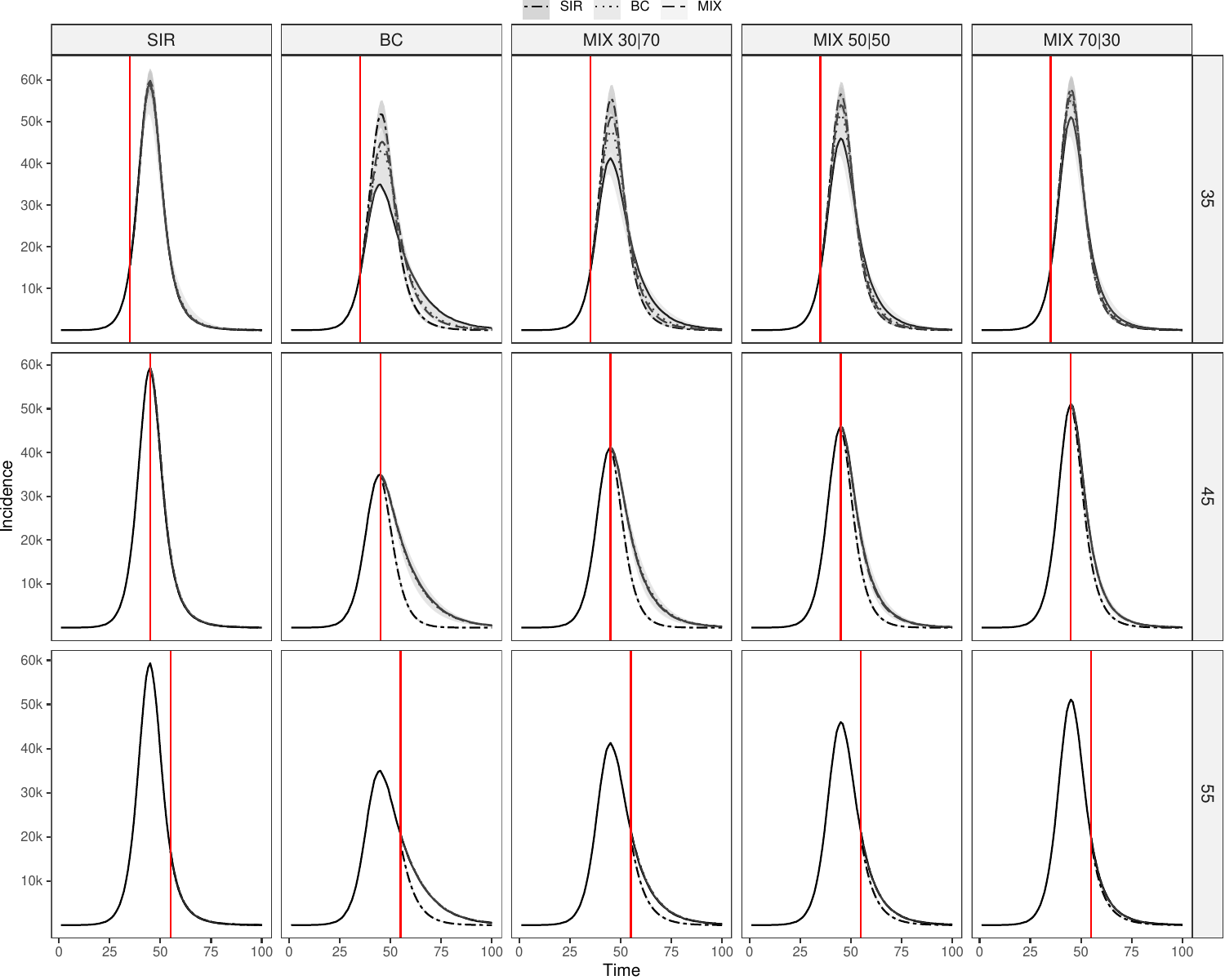}
    \caption{Posterior predictive mean (dashed, dotted, and long-dashed lines for SIR, BC, and mixture models, respectively) and 95\% credible intervals for forecasts of future incidence compared to the true values (solid line) for a randomly selected simulation. Results were generated considering different cutoffs at days 35, 45, and 55. BC and MIX curves are overlapping.}
    \label{fig:Hill2pFor}
\end{figure}

\newpage
\section{Application} \label{SM:app}

\subsection{Covid datasets - New York City}

\begin{table}[H] \centering
\caption{Convergence criteria - New York City. \label{tab1}}%
\begin{tabular}{cccccccccccc}
\toprule
\multirow{2}{*}{Model fitted} & \multirow{2}{*}{Parameter} & \multicolumn{5}{c}{Wave 1} & \multicolumn{5}{c}{Wave 2} \\
\cmidrule{3-12}
 &  & $\hat{R}$ & $\hat{R}$Upper & ESS & ESS & ESS & $\hat{R}$ & $\hat{R}$Upper & ESS & ESS & ESS\\
\midrule
\multirow{2}{*}{SIR} & $\gamma$ & 1.003 & 1.008 & 1407 & 1456 & 1322 & 1.001 & 1.005 & 694 & 706 & 634\\
 & $\beta$ & 1.003 & 1.010 & 1387 & 1367 & 1438 & 1.002 & 1.006 & 719 & 708 & 634\\
\addlinespace
\multirow{3}{*}{BC-Power} & k & 1.000 & 1.001 & 5222 & 5255 & 4805 & 1.142 & 1.159 & 447 & 676 & 3241\\
 & $\gamma$ & 1.003 & 1.010 & 1128 & 904 & 998 & 1.002 & 1.004 & 535 & 742 & 592\\
 & $\beta$ & 1.003 & 1.010 & 1276 & 1069 & 1188 & 1.002 & 1.004 & 569 & 733 & 641\\
\addlinespace
\multirow{4}{*}{MIX-Power} & k & 1.003 & 1.011 & 4508 & 3757 & 3937 & 1.141 & 1.148 & 3018 & 4896 & 1226\\
 & $\omega$ & 1.003 & 1.010 & 1767 & 1741 & 1596 & 1.000 & 1.000 & 10000 & 9685 & 9535\\
 & $\gamma$ & 1.007 & 1.022 & 497 & 496 & 552 & 1.001 & 1.002 & 661 & 709 & 597\\
 & $\beta$ & 1.005 & 1.019 & 1552 & 1477 & 1744 & 1.001 & 1.001 & 680 & 740 & 618\\
\addlinespace
\multirow{4}{*}{BC-Threshold} & H & 1.000 & 1.000 & 10671 & 10000 & 9925 & 1.000 & 1.000 & 10542 & 9006 & 9225\\
 & $\delta$ & 1.001 & 1.002 & 2920 & 2790 & 2068 & 1.001 & 1.004 & 5944 & 6604 & 5318\\
 & $\gamma$ & 1.001 & 1.002 & 691 & 716 & 707 & 1.002 & 1.006 & 646 & 664 & 634\\
 & $\beta$ & 1.000 & 1.000 & 1582 & 1537 & 1716 & 1.002 & 1.006 & 643 & 666 & 622\\
\addlinespace
\multirow{5}{*}{MIX-Threshold} & H & 1.000 & 1.000 & 10000 & 10000 & 10000 & 1.000 & 1.000 & 9341 & 9212 & 9342\\
 & $\delta$ & 1.001 & 1.002 & 1171 & 1158 & 1094 & 1.000 & 1.001 & 8133 & 7393 & 5457\\
 & $\omega$ & 1.002 & 1.006 & 930 & 1345 & 1205 & 1.000 & 1.000 & 9244 & 9331 & 8297\\
 & $\gamma$ & 1.001 & 1.005 & 900 & 658 & 795 & 1.002 & 1.005 & 739 & 769 & 687\\
 & $\beta$ & 1.000 & 1.002 & 1853 & 1300 & 1643 & 1.002 & 1.004 & 730 & 768 & 707\\
\addlinespace
\multirow{4}{*}{BC-Hill} & $\delta$ & 1.137 & 1.277 & 218 & 97 & 144 & 1.000 & 1.001 & 8256 & 6968 & 8930\\
 & $\nu$ & 1.183 & 1.538 & 41 & 88 & 52 & 1.001 & 1.002 & 6584 & 2181 & 7516\\
 & $\gamma$ & 1.019 & 1.066 & 518 & 452 & 431 & 1.005 & 1.018 & 724 & 707 & 699\\
 & $\beta$ & 1.257 & 1.825 & 53 & 73 & 46 & 1.005 & 1.017 & 706 & 722 & 696\\
\addlinespace
\multirow{4}{*}{MIX-Hill} & $\delta$ & 1.001 & 1.003 & 1072 & 1438 & 1072 & 1.000 & 1.000 & 8580 & 8030 & 4950\\
 & $\nu$ & 1.000 & 1.001 & 1630 & 2462 & 1686 & 1.000 & 1.000 & 7030 & 4864 & 1744\\
 & $\omega$ & 1.001 & 1.004 & 859 & 1150 & 931 & 1.000 & 1.000 & 10000 & 10000 & 10000\\
 & $\gamma$ & 1.001 & 1.003 & 539 & 476 & 481 & 1.002 & 1.004 & 692 & 738 & 626\\
 & $\beta$ & 1.001 & 1.002 & 816 & 1027 & 834 & 1.002 & 1.003 & 687 & 743 & 627\\
\bottomrule
\end{tabular}
\begin{tablenotes}%
\item Values of gr below 1.1 indicate satisfactory convergence.
\item Large values of ESS indicate the chain has effectively explored the posterior well, which means a good sign of convergence
\end{tablenotes}
\end{table}

\begin{table}[H] \centering
\caption{Posterior mean, 95\% credible intervals, and coverage rate for all parameters - New York City. \label{tab1}}%
\begin{tabular}{cccccccc}
\toprule
\multirow{2}{*}{Model fitted} & \multirow{2}{*}{Parameter} & \multicolumn{3}{c}{Wave 1} & \multicolumn{3}{c}{Wave 2} \\
\cmidrule{3-8}
 &  & Mean & Low & Upp & Mean & Low & Upp \\
\midrule
\multirow{2}{*}{SIR} & $\gamma$ & 2.961 & 2.697 & 3.258 & 0.880 & 0.817 & 0.951\\
 & $\beta$ & 0.961 & 0.945 & 0.976 & 0.603 & 0.576 & 0.633\\
 \addlinespace
\multirow{3}{*}{BC-Power} & k & 0.003 & 0.003 & 0.003 & 0.024 & 0.013 & 0.052\\
 & $\gamma$ & 2.436 & 2.187 & 2.705 & 0.847 & 0.784 & 0.914\\
 & $\beta$ & 1.029 & 1.004 & 1.053 & 0.598 & 0.571 & 0.626\\
\addlinespace
\multirow{4}{*}{MIX-Power} & k & 0.000 & 0.000 & 0.000 & 0.019 & 0.004 & 0.034\\
 & $\omega$ & 0.539 & 0.511 & 0.567 & 0.485 & 0.206 & 0.767\\
 & $\gamma$ & 1.771 & 1.539 & 2.031 & 0.846 & 0.782 & 0.914\\
 & $\beta$ & 1.519 & 1.436 & 1.604 & 0.598 & 0.570 & 0.627\\
 \addlinespace
\multirow{4}{*}{BC-Threshold} & H & 0.000 & 0.000 & 0.000 & 0.000 & 0.000 & 0.000\\
 & $\delta$ & 0.304 & 0.287 & 0.321 & 0.033 & 0.026 & 0.040\\
 & $\gamma$ & 2.261 & 2.023 & 2.533 & 0.812 & 0.752 & 0.878\\
 & $\beta$ & 1.288 & 1.250 & 1.326 & 0.579 & 0.551 & 0.608\\
 \addlinespace
\multirow{5}{*}{MIX-Threshold} & H & 0.000 & 0.000 & 0.000 & 0.000 & 0.000 & 0.000\\
 & $\delta$ & 0.647 & 0.395 & 0.967 & 0.088 & 0.041 & 0.214\\
 & $\omega$ & 0.502 & 0.235 & 0.686 & 0.555 & 0.244 & 0.841\\
 & $\gamma$ & 2.265 & 2.025 & 2.534 & 0.813 & 0.754 & 0.878\\
 & $\beta$ & 1.288 & 1.249 & 1.326 & 0.580 & 0.552 & 0.608\\
 \addlinespace
\multirow{4}{*}{BC-Hill} & $\nu$ & - & - & - & 7.839 & 4.706 & 12.115\\
 & $\delta$ & - & - & - & 0.001 & 0.001 & 0.001\\
 & $\gamma$ & - & - & - & 0.830 & 0.769 & 0.895\\
 & $\beta$ & - & - & - & 0.587 & 0.559 & 0.615\\
 \addlinespace
\multirow{4}{*}{MIX-Hill} & $\nu$ & 1.170 & 0.957 & 1.397 & 7.836 & 4.774 & 11.888\\
 & $\delta$ & 0.000 & 0.000 & 0.000 & 0.001 & 0.001 & 0.001\\
 & $\omega$ & 0.463 & 0.408 & 0.512 & 0.466 & 0.196 & 0.741\\
 & $\gamma$ & 1.649 & 1.439 & 1.892 & 0.828 & 0.768 & 0.892\\
 & $\beta$ & 1.658 & 1.511 & 1.849 & 0.586 & 0.558 & 0.614\\
\bottomrule
\end{tabular}
\begin{tablenotes}%
\item 
\end{tablenotes}
\end{table}

\subsection{Covid datasets - Montreal}

\begin{table}[H] \centering
\caption{Convergence criteria - Montreal. \label{tab1}}%
\begin{tabular}{cccccccccccc}
\toprule
\multirow{2}{*}{Model fitted} & \multirow{2}{*}{Parameter} & \multicolumn{5}{c}{Wave 1} & \multicolumn{5}{c}{Wave 2} \\
\cmidrule{3-12}
 &  & $\hat{R}$ & $\hat{R}$Upper & ESS & ESS & ESS & $\hat{R}$ & $\hat{R}$Upper & ESS & ESS & ESS\\
\midrule
\multirow{2}{*}{SIR} & $\gamma$ & 1.000 & 1.002 & 1638 & 1590 & 1546 & 1.001 & 1.003 & 984 & 943 & 993\\
 & $\beta$ & 1.000 & 1.001 & 1527 & 1531 & 1573 & 1.001 & 1.002 & 982 & 944 & 1000\\
\addlinespace
\multirow{3}{*}{BC-Power} & k & 1.000 & 1.000 & 4663 & 4504 & 4758 & 1.011 & 1.016 & 3870 & 2482 & 3807\\
 & $\gamma$ & 1.001 & 1.002 & 1165 & 1291 & 1073 & 1.006 & 1.019 & 892 & 824 & 826\\
 & $\beta$ & 1.001 & 1.003 & 1293 & 1464 & 1323 & 1.005 & 1.018 & 873 & 847 & 844\\
\addlinespace
\multirow{4}{*}{MIX-Power} & k & 1.000 & 1.000 & 5720 & 5869 & 4282 & 1.283 & 1.316 & 8960 & 1694 & 4126\\
 & $\omega$ & 1.000 & 1.000 & 2805 & 2773 & 2379 & 1.000 & 1.000 & 10000 & 10000 & 10000\\
 & $\gamma$ & 1.001 & 1.003 & 1249 & 1188 & 1040 & 1.011 & 1.035 & 812 & 807 & 762\\
 & $\beta$ & 1.001 & 1.001 & 2394 & 2382 & 2149 & 1.010 & 1.034 & 898 & 875 & 805\\
\addlinespace
\multirow{4}{*}{BC-Threshold} & H & 1.000 & 1.002 & 3123 & 3218 & 3276 & 1.002 & 1.003 & 5622 & 6864 & 7865\\
 & $\delta$ & 1.000 & 1.001 & 3088 & 3077 & 2833 & 1.000 & 1.001 & 4880 & 4634 & 5822\\
 & $\gamma$ & 1.002 & 1.007 & 1126 & 1195 & 1294 & 1.000 & 1.001 & 878 & 930 & 843\\
 & $\beta$ & 1.001 & 1.005 & 1449 & 1513 & 1614 & 1.000 & 1.001 & 873 & 917 & 855\\
\addlinespace
\multirow{5}{*}{MIX-Threshold} & H & 1.001 & 1.001 & 3356 & 3223 & 3489 & 1.001 & 1.001 & 7433 & 8864 & 8349\\
 & $\delta$ & 1.001 & 1.003 & 4039 & 6056 & 3861 & 1.000 & 1.000 & 8504 & 8513 & 7858\\
 & $\omega$ & 1.000 & 1.001 & 5716 & 6887 & 5520 & 1.000 & 1.000 & 10000 & 10000 & 10000\\
 & $\gamma$ & 1.004 & 1.014 & 1446 & 1264 & 1081 & 1.002 & 1.007 & 917 & 794 & 871\\
 & $\beta$ & 1.002 & 1.007 & 1899 & 1686 & 1383 & 1.002 & 1.007 & 889 & 802 & 874\\
\addlinespace
\multirow{4}{*}{BC-Hill} & $\delta$ & 1.050 & 1.053 & 1547 & 1798 & 2303 & 1.001 & 1.002 & 1696 & 9265 & 1645\\
 & $\nu$ & 1.001 & 1.003 & 1286 & 1026 & 1323 & 1.659 & 3.394 & 209 & 0 & 266\\
 & $\gamma$ & 1.004 & 1.012 & 900 & 831 & 1137 & 1.719 & 2.748 & 764 & 866 & 780\\
 & $\beta$ & 1.008 & 1.016 & 630 & 603 & 788 & 3.604 & 7.324 & 139 & 869 & 149\\
\addlinespace
\multirow{4}{*}{MIX-Hill} & $\delta$ & 1.129 & 1.131 & 1754 & 2841 & 2053 & 1.903 & 3.916 & 2 & 2362 & 5516\\
 & $\nu$ & 1.000 & 1.001 & 3885 & 3621 & 4563 & 1.618 & 3.114 & 3 & 192 & 961\\
 & $\omega$ & 1.001 & 1.002 & 1305 & 1258 & 1625 & 1.385 & 2.404 & 1 & 1591 & 1786\\
 & $\gamma$ & 1.002 & 1.004 & 1164 & 1050 & 1124 & 1.278 & 1.751 & 968 & 782 & 813\\
 & $\beta$ & 1.002 & 1.005 & 1217 & 1042 & 1429 & 2.085 & 3.752 & 967 & 412 & 468\\
\bottomrule
\end{tabular}
\begin{tablenotes}%
\item Values of gr below 1.1 indicate satisfactory convergence.
\item Large values of ESS indicate the chain has effectively explored the posterior well, which means a good sign of convergence
\end{tablenotes}
\end{table}

\begin{table}[!t] \centering
\caption{Posterior mean, 95\% credible intervals, and coverage rate for all parameters - Montreal.  \label{tab1}}%
\begin{tabular}{cccccccc}
\toprule
\multirow{2}{*}{Model fitted} & \multirow{2}{*}{Parameter} & \multicolumn{3}{c}{Wave 1} & \multicolumn{3}{c}{Wave 2} \\
\cmidrule{3-8}
 &  & Mean & Low & Upp & Mean & Low & Upp \\
\midrule
\multirow{2}{*}{SIR} & $\gamma$ & 1.332 & 1.129 & 1.561 & 0.788 & 0.670 & 0.933\\
 & $\beta$ & 0.742 & 0.682 & 0.799 & 0.561 & 0.503 & 0.624\\
 \addlinespace
\multirow{3}{*}{BC-Power} & k & 0.001 & 0.001 & 0.002 & 0.007 & 0.005 & 0.014\\
 & $\gamma$ & 1.131 & 0.937 & 1.365 & 0.686 & 0.575 & 0.821\\
 & $\beta$ & 0.766 & 0.696 & 0.839 & 0.533 & 0.475 & 0.597\\
\addlinespace
\multirow{4}{*}{MIX-Power} & k & 0.000 & 0.000 & 0.000 & 0.004 & 0.001 & 0.008\\
 & $\omega$ & 0.572 & 0.495 & 0.653 & 0.487 & 0.207 & 0.772\\
 & $\gamma$ & 0.882 & 0.714 & 1.081 & 0.687 & 0.575 & 0.819\\
 & $\beta$ & 1.007 & 0.865 & 1.169 & 0.533 & 0.475 & 0.596\\
 \addlinespace
\multirow{4}{*}{BC-Threshold} & H & 0.000 & 0.000 & 0.000 & 0.000 & 0.000 & 0.000\\
 & $\delta$ & 0.228 & 0.176 & 0.283 & 0.062 & 0.038 & 0.085\\
 & $\gamma$ & 1.047 & 0.864 & 1.264 & 0.690 & 0.584 & 0.819\\
 & $\beta$ & 0.833 & 0.749 & 0.924 & 0.518 & 0.462 & 0.582\\
 \addlinespace
\multirow{5}{*}{MIX-Threshold} & H & 0.000 & 0.000 & 0.000 & 0.000 & 0.000 & 0.000\\
 & $\delta$ & 0.525 & 0.278 & 0.922 & 0.164 & 0.068 & 0.406\\
 & $\omega$ & 0.528 & 0.239 & 0.758 & 0.556 & 0.248 & 0.841\\
 & $\gamma$ & 1.054 & 0.869 & 1.272 & 0.689 & 0.582 & 0.819\\
 & $\beta$ & 0.834 & 0.751 & 0.927 & 0.518 & 0.461 & 0.582\\
 \addlinespace
\multirow{4}{*}{BC-Hill} & $\nu$ & 0.206 & 0.141 & 0.302 & - & -  & - \\
 & $\delta$ & 0.001 & 0.000 & 0.004 & - & -  & - \\
 & $\gamma$ & 0.988 & 0.797 & 1.215 & - & -  & - \\
 & $\beta$ & 1.182 & 0.912 & 1.597 & - & -  & - \\
 \addlinespace
\multirow{4}{*}{MIX-Hill} & $\nu$ & 1.057 & 0.594 & 1.673 & 0.147 & 0.000 & 0.536\\
 & $\delta$ & 0.000 & 0.000 & 0.000 & 0.537 & 0.008 & 0.939\\
 & $\omega$ & 0.487 & 0.357 & 0.609 & 0.413 & 0.167 & 0.607\\
 & $\gamma$ & 0.866 & 0.698 & 1.067 & 0.741 & 0.614 & 0.894\\
 & $\beta$ & 1.125 & 0.908 & 1.442 & 0.663 & 0.525 & 0.818\\
\bottomrule
\end{tabular}
\begin{tablenotes}%
\item 
\end{tablenotes}
\end{table}

\end{document}